\documentclass[11pt]{article}
\usepackage{bm}
\usepackage[font=small]{caption}
\RequirePackage{amsthm,amsmath,amsfonts,amssymb}
\usepackage{omega}
\usepackage[a4paper,left=2.5cm,right=2.5cm,top=2.5cm,bottom=2.5cm]{geometry}

\usepackage{bm}
\usepackage{natbib}
\usepackage[ruled]{algorithm2e}
\usepackage[dvipsnames]{xcolor}
\usepackage[colorlinks = true,
            linkcolor = NavyBlue,
            urlcolor  = NavyBlue,
            citecolor = NavyBlue,
            anchorcolor = NavyBlue]{hyperref}

\usepackage{url}
\usepackage[graphicx]{realboxes}
\usepackage{enumerate}
\usepackage{comment}
\usepackage{subfigure}
\usepackage[title]{appendix}
\usepackage{chngcntr}
\RequirePackage{subfigure}
\RequirePackage{algorithmic}

\author{}
\usepackage{setspace}

\newcommand{\T}{\top}
\newcommand{\RR}{\mathbb{R}}
\newcommand{\ee}{\end{aligned} \end{equation}}
\newcommand{\eq}{\end{quote}}

\newcommand{\ep}{\end{parts}}

\newcommand{\bqp}{\begin{quote}\begin{parts}}

\newcommand{\epq}{\end{parts}\end{quote}}

\DeclareMathOperator*{\argmin}{argmin}
\newcommand{\Rom}[1]{\text{\uppercase\expandafter{\romannumeral #1\relax}}}
\newcommand{\bee}{\begin{equation}\begin{aligned}}

\newcommand{\emm}{\end{bmatrix}}

\numberwithin{equation}{section}
\newcommand{\vertiii}[1]{{\vert\kern-0.25ex\vert\kern-0.25ex\vert #1 
    \vert\kern-0.25ex\vert\kern-0.25ex\vert}}
    
\newcommand\ma{\mathbf{A}}
\newcommand\mb{\mathbf{B}}

\newcommand\mi{\mathbf{I}}

\newcommand\mr{\mathbf{R}}

\newcommand\mx{\mathbf{X}}

\newcommand\mz{\mathbf{Z}}
\newcommand\mw{\mathbf{W}}

\newcommand\mm{\mathbf{M}}

\newcommand\mpp{\mathbf{P}}
\newcommand\mo{\mathbf{O}}
\newcommand\my{\mathbf{Y}}

\newcommand{\SVD}{\operatorname{SVD}}

\newcommand{\bea}{\begin{eqnarray}}
	\newcommand{\eea}{\end{eqnarray}}
\newcommand{\beas}{\begin{eqnarray*}}
	\newcommand{\eeas}{\end{eqnarray*}}
	
\newcommand{\red}{\color{black}}

\renewcommand{\(}{\left(}
\renewcommand{\)}{\right)}

\title{A Functional Tensor Model for Dynamic Multilayer Networks with Common Invariant Subspaces and the RKHS Estimation}
\begin{document}

\author{Runshi Tang\thanks{The first two authors contributed equally}
\thanks{Department of Statistics, University of Wisconsin-Madison}
\and Runbing Zheng\footnotemark[1]
\thanks{Department of Applied Mathematics and Statistics, Johns Hopkins University}
\and Anru R. Zhang
\thanks{Department of Biostatistics \& Bioinformatics and Department of Computer Science, Duke University}
\and Carey E. Priebe\footnotemark[3]}

\date{}

\maketitle

\begin{abstract}
Dynamic multilayer networks are frequently used to describe the structure and temporal evolution of multiple relationships among common entities, with applications in fields such as sociology, economics, and neuroscience. However, exploration of analytical methods for these complex data structures remains limited.
We propose a functional tensor-based model for dynamic multilayer networks, with the key feature of capturing the shared structure among common vertices across all layers, while simultaneously accommodating smoothly varying temporal dynamics and layer-specific heterogeneity.
The proposed model and its embeddings can be applied to various downstream network inference tasks, including dimensionality reduction, vertex community detection, analysis of network evolution periodicity, visualization of dynamic network evolution patterns, and evaluation of inter-layer similarity.
We provide an estimation algorithm based on functional tensor Tucker decomposition and the reproducing kernel Hilbert space framework, with an effective initialization strategy to improve computational efficiency. 
The estimation procedure can be extended to address more generalized functional tensor problems, as well as to handle missing data or unaligned observations.
We validate our method on simulated data and two real-world cases: the dynamic Citi Bike trip network and an international food trade dynamic multilayer network, with each layer corresponding to a different product.

\end{abstract}

\noindent%
\vfill

\vfill

\begin{sloppypar}

\counterwithout{figure}{section}
\counterwithout{theorem}{section}
\counterwithout{definition}{section}

\theoremstyle{definition}
\newtheorem{rema}{Remark}

\newtheorem{propositi}{Proposition}

\section{Introduction}

%{\color{red} A few papers I feel related that we could cite: 
%1. Global and individualized community detection in inhomogeneous multilayer networks; 
%2. Community detection with contextual multilayer networks. 
%3. Consistent community detection in multi-layer network data

%A review of dynamic network models with latent variables

%Spectral clustering via adaptive layer aggregation for multi-layer networks

%}

Networks are powerful for representing relationships or connections among a collection of entities, and dynamic networks can further describe how the structure of these relationships evolves over time, with significant applications in fields such as sociology \citep{zuzul2025dynamic, rand2011dynamic}, economics \citep{karim2022determining, wanke2019dynamic}, and neuroscience \citep{eschbach2020recurrent, ju2020dynamic}.
A single dynamic network describes the evolution of a specific relationship among entities, while a dynamic multilayer network captures the co-evolution of multiple relationships among shared entities.
For instance, the temporal changes in trade for a specific commodity between countries can be represented as a single dynamic trade network, while incorporating different commodities as separate layers results in a dynamic multilayer trade network \citep{barigozzi2010multinetwork}. 
Similarly, the temporal evolution of brain region connectivity within a specific frequency band can be represented as a single dynamic brain network, with different frequency bands treated as separate layers giving rise to a dynamic multilayer brain network \citep{buldu2018frequency}.

Dynamic multilayer networks, consisting of networks observed at different time points and across different layers, can be regarded as a special type of multiple networks, whereas general multiple networks typically assume that there is no specific order among the networks and that the networks share certain common structure. % due to the shared vertices. 
One prominent example is the multilayer stochastic block model (SBM) \citep{holland1983stochastic, han2015consistent, paul2020spectral, lei2020consistent, lei2023bias, huang2023spectral, lei2024computational}, which assumes that vertices share community assignments across different layers while allowing for layer-specific block probabilities.
\cite{chen2022global} considers its variant allowing each layer to deviate from the global community assignment with a certain probability, and \cite{ma2023community} considers a variant that incorporates an observed covariate matrix encoding the same common community structure.
Other examples include multilayer eigenscaling models \citep{nielsen2018multiple, wang2019joint, draves2020bias, weylandt2022multivariate}, which assume that the probability matrices of the networks share common subspaces and constrain the low-dimensional heterogeneous structure to be diagonal,
and the common subspace independent edge (COSIE) model \citep{arroyo2021inference}, which also assumes that the probability matrices share common subspaces but imposes no specific constraint on the low-dimensional heterogeneous structure.
Although substantial research has been conducted on general multiple networks, studies on dynamic networks, especially dynamic multilayer networks, remain relatively limited.

A challenge in modeling dynamic networks or dynamic multilayer networks lies in characterizing the temporal evolution, as their observed networks, unlike general multiple networks, have a well-defined temporal order that can be leveraged.
Most existing models that characterize this temporal order can be broadly divided into two categories: (1) dependent models, which assume that the network at the next time point depends on its states at previous time points following a consistent evolutionary pattern, and (2) smooth models, which assume that the network evolves smoothly over time.
Dependent models include the latent space models for dynamic networks proposed in \cite{sarkar2005dynamic} and \cite{sewell2015latent}, which assume connection probabilities are determined by distances between vertex latent positions, which evolve according to Gaussian random walks. %the connection probabilities between vertices are determined by the distance between their latent positions, and that the latent positions for each vertex evolve according to a Gaussian random walk. %, with each step following a Gaussian distribution centered on the previous step.
\cite{heaukulani2013dynamic} uses generalized inner products of binary vertex latent feature vectors, which evolve via Bernoulli distributions with transition probabilities depending on a weighted sum of the same feature across neighboring vertices from the previous step. 
%assumes that connection probabilities between vertices are determined by a generalized inner product of their binary latent feature vectors, and that each latent feature evolves according to a Bernoulli distribution with transition probabilities depending on a weighted sum of the same feature across neighboring vertices from the previous step.
\cite{mazzarisi2020dynamic} assumes links are formed either by copying from the previous time step with a constant probability, or according to connection probabilities determined by vertex latent features, where features follow Gaussian autoregressive processes.
%\cite{mazzarisi2020dynamic} assumes that links between vertices are formed either by copying the presence or absence of the link from the previous time step with a constant probability, or according to probabilities determined by their latent features, and that the latent feature for each vertex evolves over time according to a Gaussian autoregressive process. % dependent on the previous time step.
%\cite{ishiguro2010dynamic} assumes that connection probabilities between vertices are determined by the interaction probabilities between their respective cluster assignments, and that the cluster assignment for each vertex evolves over time according to a hierarchical Dirichlet process with transition probabilities that encourage persistence in the cluster assignment from the previous step.
{\color{black} \cite{jiang2023autoregressive} proposes an autoregressive model where edges retain their previous states or transition between states according to edge-specific probabilities; when applied to stochastic block models, communities are characterized by community-level transition probabilities.
\cite{jiang2025two} develops a two-way heterogeneity model for autoregressive dynamic networks that assigns each node two parameters, one characterizing its propensity to form ties and the other characterizing its tendency to retain existing ties.}
Building on the SBM, several models focus on evolving community structures.
\cite{ishiguro2010dynamic} uses hierarchical Dirichlet processes for cluster assignments with persistence-encouraging transition probabilities,
%assumes that the cluster assignment for each vertex evolves over time according to a hierarchical Dirichlet process, with transition probabilities that encourage persistence in the cluster assignment from the previous step, 
while \cite{matias2017statistical} assumes vertex communities follow Markov chains.
%assumes vertex community assignments follow a Markov chain.
\cite{daniel2023bayesian} combines cluster-specific interactions with latent position distances, where cluster assignments evolve via hierarchical Dirichlet processes conditional on the previous cluster assignments, and positions evolve via Gaussian autoregressive processes conditionally on current cluster assignments.
%\cite{daniel2023bayesian} assumes that connection probabilities between vertices are determined by their cluster-specific interaction patterns and the distance of their latent positions, and assumes for each vertex, the cluster assignment evolves according to a hierarchical Dirichlet process conditional on the previous cluster assignment, and the latent position evolves according to a Gaussian autoregressive process conditional on the current cluster assignment. % and previous latent position.
See the review in \cite{kim2018review} for more dependent models of dynamic networks.
\cite{loyal2023eigenmodel} and \cite{lan2025tensor} consider dynamic multilayer networks.
\cite{loyal2023eigenmodel} assumes connection probabilities in each layer are determined by the generalized inner product of vertex latent positions with layer-specific kernels plus vertex layer-and-time-specific effects, with positions and effects evolving via independent Gaussian random walks.
%assumes that connection probabilities between vertices in each layer are determined by the generalized inner product of their latent positions with layer-specific kernels plus their layer-and-time-specific effects, and that the latent position and the effect for each vertex evolve over time according to independent Gaussian random walks. % centered on their previous time step.
\cite{lan2025tensor} uses time-varying tensor decomposition with a fixed node-specific feature matrix and a fixed layer-specific transition matrix, where the core tensor evolves via Gaussian autoregressive processes.
%\cite{lan2025tensor} assumes that connection probabilities between vertices are determined by %a Tucker tensor decomposition of 
%a time-varying tensor decomposition with a node-specific feature matrix and a layer-specific transition matrix, and that the core tensor evolves over time according to a Gaussian autoregressive process. %dependent on the previous time step with Gaussian noise.
{\color{black}  \cite{lopez2022dynamic} considers a dynamic multilayer SBM where community membership transitions in each layer depend on past memberships across layers. %, enabling identification of directed causal dependencies between layers.
}

Although dependent models capture the intrinsic temporal evolution patterns of networks, these models are typically suitable for networks whose dynamics are primarily driven by their own states (either observed or latent).
However, in practice, many network dynamics may be influenced by external factors. For example, temporal changes in international trade networks may result from global economic events, and temporal changes in brain networks may be driven by specific external stimuli. In such cases, dependent models may no longer be applicable, and smooth models, which do not make assumptions about the drivers of network evolution, might be more appropriate.
Smooth models allow for arbitrary, continuously varying latent processes, with estimation efficiency depending on the exploitation of the temporal smoothness of networks.
Existing smooth models for dynamic networks include the study in \cite{kreiss2019nonparametric}, which model edge formation using counting processes with time-varying intensities that depend on smooth parameter vectors. \cite{matias2018semiparametric} adopts a similar model but incorporates latent community structures.
\cite{macdonald2022latent,shi2025exploratory} assume that vertices have time-dependent functional latent positions that vary smoothly over time, and that the connection probabilities between vertices are the inner products of their respective latent positions.
\cite{wang2024representation} considers directed dynamic networks, assuming that vertices have fixed incoming latent positions and smooth time-dependent functional outgoing latent positions, with connection probabilities determined by inner products between outgoing and incoming positions, and the estimation approach encourages community structure.

In this paper, we propose a smooth model for dynamic networks motivated by the COSIE model, which is widely effective for general multiple networks, and then extend it to dynamic multilayer networks. The COSIE model is flexible enough to encompass many popular multiple network models, including the multilayer SBM and multiple eigenscaling models mentioned above, and is also tractable for inference tasks such as vertex community detection and the evaluation of similarity across networks. 
For the COSIE model, the common subspaces can be viewed as latent positions of shared vertices, and connection probabilities between vertices in each individual network are the generalized inner products of these positions with the kernel matrix defined by the heterogeneous structure.
We adopt a similar model structure using the generalized inner product with varying kernel to preserve its flexibility and tractability, and further assume that the kernel matrix evolves smoothly over time, enabling us to leverage shared local information between snapshots of the dynamic network to achieve efficient estimation of the network structure.
Our model separates the network representation into time-invariant vertex latent positions and smooth time-varying low-dimensional kernels, facilitating various inference tasks. Specifically, the latent positions can be used for vertex community detection, while the time-varying functional kernels characterize the dynamic network evolution in low-dimensional space and can be used to analyze its periodicity or further dimensionally reduced for network evolution visualization.
In addition, this approach of decomposing the varying component into low-dimensional space can provide more accurate estimation compared to models with high-dimensional varying structures \citep{macdonald2022latent,shi2025exploratory,wang2024representation} particularly when the varying component is indeed low-dimensional, such as for the dynamic SBM \citep{bhattacharyya2018spectral} where vertex community assignments remain fixed while community probabilities evolve smoothly over time.
Moreover, for undirected networks, standard inner product models \citep{macdonald2022latent,shi2025exploratory} are limited to positive semi-definite probability matrices, whereas our model with a generalized inner product has no such restriction.
When extending to dynamic multilayer networks, we assume that all layers on the same vertices share common invariant subspaces, and that each layer's kernel matrix evolves smoothly over time.
The trajectories of these layer kernels not only characterize their individual evolution patterns but also capture the relationships between layers.

We also make contributions to the estimation methodology. By reformulating the generalized inner product model with varying kernel, we obtain a functional tensor Tucker decomposition. We develop an efficient algorithm for it based on the reproducing kernel Hilbert space (RKHS) framework, and propose an effective initialization strategy to enhance computational efficiency. Our estimation procedure can actually be extended to tackle functional tensor Tucker decomposition of arbitrary order with various types of errors, as well as handling missing data and unaligned observations.

The structure of our paper is as follows. 
Section~\ref{sec:notation} introduces some notations and relevant preliminary concepts of functional tensors and RKHS.
In Section~\ref{sec:model}, we present our functional tensor-based model for a single dynamic network and extend it to dynamic multilayer networks. Section~\ref{sec:estimation} provides the estimation algorithm based on the RKHS framework. Section~\ref{sec:simu} validates the accuracy of our estimation algorithm and demonstrates the effectiveness of our method for inference tasks through simulation studies. Section~\ref{sec:real} applies our method to two real datasets: Citi Bike trip data for a single dynamic network, and international food trade data for a dynamic multilayer network.
Section~\ref{sec:extension} presents extensions of our network model and estimation algorithm, including generalizing our model for more complex inter-layer dependencies or for scenarios where layers follow a known ordering that allows for additional smoothness assumptions across layers, and extending our estimation procedure to tackle more generalized tensor decomposition problems.
Finally in Section~\ref{sec:discussion}, we discuss how our method can be applied to networks with abrupt changes or to offline learning, as well as its potential extensions to more complex data structures.
Proofs of the stated results, additional discussion, and additional experimental results are provided in the appendix.

\section{Notation and Preliminaries}\label{sec:notation}

We summarize the notation used in this paper. For any positive integer $n$, we denote by $[n]$ the set $\{1,2,\dots,n\}$.
We denote by $\mathcal{O}_d$ the set of $d\times d$ orthogonal matrices, and denote by $\mathcal{O}_{n\times d}$ the set of $n\times d$ matrices with orthonormal columns.
We use $\operatorname{SVD}_{d}(\mm)$ to denote the matrix whose columns are the $d$ leading left singular vectors of $\mm$, arranged in order.

In many applications, data naturally take the form of multi-way arrays, or \emph{tabular tensors}. An order-3 tabular tensor $\mathbf{P}$ can be viewed as a mapping  
$
[n] \times [p] \times [q] \to \mathbb{R}, (i,j,k) \mapsto \mathbf{P}_{i,j,k}.
$
However, in a different class of problems, tensor data arise as a collection of functions. For example, suppose we observe $\mathbf{P}_{i,j}^{[t]}$, representing the value of feature $j$ for subject $i$ at time $t$. A common workaround is to discretize the time domain and represent the data as a tabular tensor. Yet, this approach treats the time index as exchangeable, failing to capture the inherent sequential or continuous structure of the functional mode.

To address this limitation, the notion of a \emph{functional tensor} has been introduced. An order-$(D+1)$ functional tensor $\mathbf{P}_{i_1, \ldots, i_D}^{[t]}$ with one functional mode can be interpreted as the mapping
$
    [p_1] \times \cdots \times [p_D] \times [0,1] \to \mathbb{R}, %\\
    (i_1, \ldots, i_D, t) \mapsto \mathbf{P}_{i_1, \ldots, i_D}^{[t]}.
$
In other words, for each $t \in [0,1]$, $\mathbf{P}^{[t]} := \{\mathbf{P}_{i_1, \ldots, i_D}^{[t]}\}_{i_k \in [p_k], k \in [D]}$ is a tabular tensor, and $\mathbf{P}$ can be viewed as a tensor-valued function over time. 
%{\color{black}(Do we use this notation?) For any tensor $\mathbf{T}$ indexed by $i_1, \ldots, i_D$, we denote by $\mathbf{T}_{i_k'}$ the order-$(D-1)$ subtensor obtained by fixing index $i_k = i_k'$.}

A central task in tensor analysis is \emph{decomposition}, which seeks a low-rank approximation of the tensor. For tabular tensors, popular decomposition methods include Tucker decomposition \citep{zhang2018tensor, luo2021sharp, agterberg2024statistical}, Canonical Polyadic (CP) decomposition \citep{Hitchcock1927, hong2020generalized, tang2025revisit}, Tensor Train \citep{zhou2022optimal}, and Tensor Ring \citep{chen2020tensor, chen2024one}. Recent developments have extended these frameworks to the functional setting, including functional Tucker decomposition \citep{hanTensorFactorModel2024} and functional CP decomposition \citep{han2024guaranteed, hanCPFactorModel2024, tang2024tensor, larsen2024tensor}.

In particular, we say that an order-$(D+1)$ functional tensor $\mathbf{P} \in \RR^{p_1 \times \cdots \times p_D \times [0,1]}$, with the functional mode indexed by $t$, has \emph{Tucker rank} $(r_1, \ldots, r_D)$ if
$
\mpp^{[t]} = \mr^{[t]} \times_{k \in [D]} \mx_k, \text{for all } t \in [0,1],
$
for some functional core tensor $\mr \in \RR^{r_1 \times \cdots \times r_D \times [0,1]}$ and orthonormal factor matrices $\mx_k \in \mathcal{O}_{p_k \times r_k}$. The operator $\times_k$ represents the $k$th mode product between a tensor and a matrix with proper dimensions. For a comprehensive introduction to tensor algebra, operations, and decomposition techniques, interested readers are directed to \cite{kolda2009tensor}. 

We now review essential concepts from the theory of RKHS.  
Let $\mathcal{H}$ be a Hilbert space equipped with an inner product $\langle \cdot, \cdot \rangle_{\mathcal{H}}$ and norm $\|\cdot\|_{\mathcal{H}}$.  
Assume there exists a kernel function $\mathbb{K}: [0,1] \times [0,1] \to \mathbb{R}_+$ that is continuous, symmetric, and positive semi-definite, and satisfies the following RKHS conditions:  
(1) for any $s \in [0,1]$, $\mathbb{K}(\cdot, s) \in \mathcal{H}$;  
(2) for any $g \in \mathcal{H}$, the reproducing property holds:
$
g(t) = \langle g, \mathbb{K}(\cdot, t) \rangle_{\mathcal{H}}$ for all $t \in [0,1].
$
The map $t \mapsto \mathbb{K}(\cdot, t)$ is referred to as the \emph{feature map}.  
Using this feature map, the Representer Theorem \citep{kimeldorf1970correspondence, scholkopf2001generalized} asserts that a broad class of regularized optimization problems over $\mathcal{H}$ admits solutions expressible as finite linear combinations of feature maps:
$
f = \sum_{i=1}^{n} \theta_i \mathbb{K}(\cdot, x_i),
$
where $\{x_i\}_{i=1}^n$ are sample points and $\{\theta_i\}$ are coefficients. 
The commonly used kernels include radial kernel $\mathbb{K}(x, y) = \exp \left({-|x-y|}\right)$, 
Bernoulli kernel $\mathbb{K}(x, y)=1+k_1(x) k_1(y)+k_2(x) k_2(y)-k_4(|x-y|)$, where $k_1(x)=x-0.5, k_2(x)=\left(k_1^2(x)-1 / 12\right) / 2$, and $k_4(x)=\left(k_1^4(x)-k_1^2(x) / 2+7 / 240\right) / 24$ for any $x \in[0,1]$, 
polynomial kernel $\mathbb{K}(x, y) = ( 0.5 x  y + 1)^3$, 
and periodic kernel $\mathbb{K}(x, y) = \exp \left(-{\sin ^2( \pi|x-y| / p)}\right)$ with period $p$. 
For further background on RKHS theory and its applications in function approximation, see \cite{aronszajn1950theory, kennedy2013hilbert, scholkopf2002learning}.

\section{Model}
\label{sec:model}

%, and will extend it to dynamic multilayer networks in Section~\ref{sec:multilayer}.

\subsection{Functional tensor model for single dynamic networks}\label{sec:single}

We begin by introducing our functional tensor-based model for a single dynamic network. %For a single dynamic network, s
Suppose there are $n$ vertices observed at time points $\{t_1, t_2, \dots, t_m\}=:\mathcal{T}$, where, for simplicity and without loss of generality, we set $0 =  t_1 < t_2 <\dots < t_m = 1$. We observe $n\times n$ binary adjacency matrices $\ma^{[t]}$ %\in \{0,1\}^{n\times n}$ 
for $t\in\mathcal{T}$. The entry $\ma^{[t]}_{i,j}$ indicates the presence ($\ma^{[t]}_{i,j}=1$) or absence ($\ma^{[t]}_{i,j}=0$) of a directed edge from vertex $i$ to vertex $j$ at time $t$.
This paper primarily considers directed networks, so $\ma^{[t]}$ can be asymmetric. The adaptation of our model and estimation algorithm to undirected networks is discussed in Remark~\ref{rmk:symmetric}. 

Our model for dynamic networks is motivated by a widely used model for multiple heterogeneous networks with shared vertices, the COSIE model \citep{arroyo2021inference}. In directed COSIE \citep{zheng2022limit}, the probability matrix of the $s$th network is given by $\mathbf{P}_s = \mathbf{X} \mathbf{R}_s \mathbf{Y}^\top$, where $\mathbf{X}, \mathbf{Y} \in \mathbb{R}^{n \times d}$ are orthonormal matrices representing low-dimensional common subspaces that characterize the shared vertices, and $\mathbf{R}_s \in \mathbb{R}^{d \times d}$ captures the heterogeneity across networks.
The COSIE model is flexible to encompass many multiple network models, including the multilayer SBM \citep{holland1983stochastic, han2015consistent, paul2020spectral, lei2020consistent, lei2023bias, huang2023spectral, lei2024computational} where vertices share community assignments, %while allowing layer-specific block probability, 
and multiple eigenscaling models \citep{nielsen2018multiple, wang2019joint, draves2020bias, weylandt2022multivariate} where each $\mathbf{R}_s$ is diagonal,
and is also tractable for performing inference tasks, such as vertex community detection and network similarity evaluation.

Dynamic networks are a special type of multiple networks with a well-defined temporal order among observed networks, and a natural expectation is that the network structure evolves continuously over time. 
Our model is designed to retain the flexibility and tractability of the COSIE model, while also leveraging dynamic-specific structures by assuming that the heterogeneous matrix $\mathbf{R}^{[t]}$ varies continuously and smoothly over time $t$. Recall that, as mentioned earlier, we set $t\in[0,1]$ without loss of generality.
Building on a functional tensor framework \citep{han2024guaranteed, hanTensorFactorModel2024, tang2024tensor, larsen2024tensor}, we propose the functional tensor dynamic network (FTDN) model as follows.

\begin{definition}[Functional tensor dynamic network]\label{def_ftdn}
Let $\mx$ and $\my$ be $n\times d$ matrices with orthonormal columns, and let $\mr^{[t]}: [0,1] \rightarrow \RR^{d \times d}, t \mapsto \mr^{[t]}$ be a smooth, matrix-valued function.
We say that the random adjacency matrices $\{\ma^{[t]}\}_{t\in\mathcal{T}\subseteq[0,1]}$ are jointly distributed according to the functional tensor dynamic network model with parameters $(\mx,\my,\mr^{[t]})$ if the observed edges follow independent Bernoulli distributions
$
\mathbf{A}^{[t]}_{ij} \sim \mathrm{Bernoulli}(\mathbf{P}^{[t]}_{i,j}),
$ 
where $\mathbf{P}^{[t]}_{ij}\in(0,1)$ denotes the edge probability from vertex $i$ to vertex $j$ at time $t$, defined by
$$
\operatorname{logit} \left(\mpp^{[t]}_{i,j}\right)= \tilde\mpp^{[t]}_{i,j}
\text{, with }
\tilde\mpp^{[t]} = \mx \mr^{[t]} \my^\top,
$$
where the logit function is defined as $\operatorname{logit}(p)=\ln \frac{p}{1-p}$ with inverse $\operatorname{logit}^{-1}(\tilde p)=\frac{\exp(\tilde p)}{1+\exp(\tilde p)}$.
\end{definition}

Here $\mathbf{X}, \mathbf{Y} \in \mathbb{R}^{n \times d}$ are time-invariant structure of the dynamic network. Each row, $\mathbf{x}_i$ and $\mathbf{y}_j$, corresponds to the outgoing and incoming fixed embeddings of vertices $i$ and $j$, respectively, in a $d$-dimensional latent space. The edge probability is given by $\tilde{\mathbf{P}}^{[t]}_{i,j} = \mathbf{x}_i^\top \mathbf{R}^{[t]} \mathbf{y}_j$. $\mathbf{R}^{[t]} \in \mathbb{R}^{d \times d}$ is a time-varying matrix that captures the evolving interactions between latent dimensions.

Compared to methods that treat dynamic networks observed at different time points as time-agnostic multiple networks, such functional model, by leveraging both the inherent time order and temporal smoothness, has the potential to yield more accurate estimates; see Section~\ref{sec:dynamic SBM} for simulation comparisons.

The FTDN model can encompass, for example, the dynamic SBM \citep{bhattacharyya2018spectral}, where the vertex community assignments remain fixed while the community probabilities evolve smoothly over time. More specifically, suppose each vertex $i$ is assigned to a community, and define the binary community assignment matrix $\mathbf{Z} \in \{0,1\}^{n \times d}$ such that $\mathbf{Z}_{ik} = 1$ if and only if vertex $i$ belongs to community $k$. Let $\mathbf{B}^{[t]} \in (0,1)^{d \times d}$ be a smooth, time-varying matrix that specifies the probabilities of edges within and between the $d$ communities at time $t$. Then the edge probability matrix is given by $\mathbf{P}^{[t]} = \mathbf{Z} \mathbf{B}^{[t]} \mathbf{Z}^\top$, which is a special case of the FTDN model with $\mathbf{X} = \mathbf{Y} = \mathbf{Z}(\mathbf{Z}^\top \mathbf{Z})^{-1/2}$, and $\mathbf{R}^{[t]} = (\mathbf{Z}^\top \mathbf{Z})^{1/2} \operatorname{logit}(\mathbf{B}^{[t]})(\mathbf{Z}^\top \mathbf{Z})^{1/2}$.
Furthermore, the FTDN model can also accommodate scenarios in which vertex community assignments also evolve over time when these changes occur smoothly. 
By representing the community structure $\mb$ using the finest partition across time, the variation in connectivity can be captured through a smooth evolution of the matrix $\mathbf{R}^{[t]}$.
{\color{black}
For example, consider a network with two communities where a subset of vertices gradually transitions from Community 1 to Community 2. Using the finest partition across time, we represent this with 3 communities: those remaining in Community 1 throughout, those remaining in Community 2 throughout, and the transitioning group (Community 3). In the corresponding $3 \times 3$ block probability matrix $\mathbf{B}^{[t]}$, Community 3 initially behaves like Community 1 and eventually transitions to behave like Community 2. The matrix $\mathbf{B}^{[t]}$ evolves smoothly over time, corresponding to smooth variation of $\mathbf{R}^{[t]}$ in the FTDN representation.
See Appendix~\ref{app:add discussion for FTDN} for more examples and detailed discussion.

}

\subsection{Multilayer functional tensor dynamic networks}\label{sec:multilayer}

Building on the FTDN model for single dynamic networks in Definition~\ref{def_ftdn}, we consider a multilayer extension.
For a dynamic $K$-layer network with a common set of $n$ vertices, we observe $n\times n$ binary adjacency matrices $\ma^{[t]}_s$ %\in \{0,1\}^{n\times n}$
 at time points $\mathcal{T}:=\{t_1, t_2, \dots, t_m\}$ where $0 =  t_1 < t_2 <\dots < t_m = 1$, for all layers $s\in[K]$. 
Each matrix $\ma^{[t]}_s$ corresponds to layer $s$ at time $t$, and the entry $\ma^{[t]}_{s,i,j}$ indicates the presence ($\ma^{[t]}_{s,i,j}=1$) or absence ($\ma^{[t]}_{s,i,j}=0$) of a directed edge from vertex $i$ to vertex $j$.

Motivated again by the COSIE model and the fact that vertices are shared across layers, we assume that all layers share common invariant subspaces. For each layer $s$, we still assume that the heterogeneous matrix $\mathbf{R}^{[t]}_s$ is a smooth function of time $t$, and propose the following multilayer functional tensor dynamic network (MFTDN) model.

\begin{definition}[Multilayer functional tensor dynamic network]\label{def_multilayer_ftdn}
Let $\mx$ and $\my$ be $n\times d$ matrices with orthonormal columns, and let $\mr^{[t]}_s: [K]\times [0,1] \rightarrow \RR^{d \times d},\ (s, t) \mapsto \mr_s^{[t]}$ be a  matrix-valued function that is smooth with respect to $t$.
We say that the random adjacency matrices $\{\ma^{[t]}_s\}_{s\in[K],t\in\mathcal{T}\subseteq[0,1]}$ are jointly distributed according to the multilayer functional tensor dynamic network model with $(\mx,\my,\mr^{[t]}_s)$ if the observed edges follow independent Bernoulli distributions
$
\mathbf{A}^{[t]}_{s,i,j} \sim \mathrm{Bernoulli}(\mathbf{P}^{[t]}_{s,i,j}),
$
where $\mathbf{P}^{[t]}_{s,i,j}\in(0,1)$ denotes the edge probability from vertex $i$ to vertex $j$ in layer $s$ at time $t$, defined by
$$
\operatorname{logit}(\mpp^{[t]}_{s,i,j})= \tilde\mpp^{[t]}_{s,i,j},
\text{ with }
\tilde\mpp_s^{[t]} = \mx \mr_s^{[t]} \my^\top.
$$
\end{definition}

$\mx,\my$ characterize the vertex features shared across layers, and $\mr^{[t]}_s$ captures the layer-specific differences and their smooth temporal evolution.
For notational simplicity, we may abbreviate the function $\mathbf{R}^{[t]}_s$ as $\mathbf{R}^{[t]}$ or $\mathbf{R}$ when the context is clear.

{\color{black}
Similar to how the FTDN model encompasses the dynamic SBM in Section~\ref{sec:single}, the MFTDN model can also accommodate dynamic multilayer SBMs. This includes not only the case where all layers share the same community assignment, but also cases where community assignments differ across layers. In the latter scenario, the distinct layer-specific structures can be captured by increasing the embedding dimension $d$ appropriately. 
See Appendix~\ref{app:add discussion for MFTDN} for more discussion.

%For example, if $2$ layers each have $2$ communities but potentially with different assignments, their union can be represented by an MFTDN with $d$ at most {\color{black}$4$}.

The MFTDN model essentially extends the COSIE model by incorporating a smoothness assumption on $\mathbf{R}_s^{[t]}$ over time $t$ (which implies smoothness of the probability matrices $\mathbf{P}_s^{[t]}$ over time $t$). For many real-world dynamic networks, we believe such a smoothness assumption is reasonable. When this smoothness assumption holds, the MFTDN model should be able to achieve more accurate estimates than the COSIE model by leveraging this additional temporal smoothness information. And our simulation studies in Section~\ref{sec:simu} empirically confirm that MFTDN achieves superior estimation accuracy compared to COSIE and other baseline models that do not exploit smoothness.

}

Note that here, for ease of exposition, we assume that for all time points in $\mathcal{T}$, we observe the presence or absence of all edges across all layers. Both the model and the estimation procedure can be extended to handle cases with missing or even unaligned observations; see Section~\ref{sec:missing}.
Additionally, notice that in MFTDN, $\mpp^{[t]}_s$ and $\mr^{[t]}_s$ are smooth functions of time $t$, with no specific ordering across layers $s$. An extension with ordered layers is discussed in Section~\ref{sec:model with smooth s}. And a generalized model with more complex layer interactions is provided in Section~\ref{sec:model with layer interaction}.

\subsection{Identifiability}\label{sec:identifiability}

Note that for any $\mw_\mx, \mw_\my \in \mathcal{O}_d$, the MFTDN with parameters $(\mx\mw_\mx, \my\mw_\my, \mw_\mx^{\top} \mr^{[t]}_s \mw_\my)$ yields the same trajectory of probability matrices $\{\mpp^{[t]}_s\}_{t \in [0,1],\ s \in [K]}$ as the MFTDN with $(\mx, \my, \mr^{[t]}_s)$, since
$
\tilde\mpp^{[t]}_s = \mx \mr^{[t]}_s \my^\top
= (\mx\mw_\mx)(\mw_\mx^\top \mr^{[t]}_s \mw_\my)(\my\mw_\my)^\top.
$
%Thus, the MFTDN model is non-identifiable up to orthogonal transformations.
Such non-identifiability is unavoidable in many latent space models, including COSIE, the random dot product graph (RDPG) \citep{athreya2018statistical}, the generalized random dot product graph (GRDPG) \citep{rubin2022statistical}, and the latent process model for functional network \citep{macdonald2022latent}.
Nevertheless, as shown in Proposition~\ref{prop:identifiability}, the MFTDN model is identifiable up to orthogonal transformations under mild conditions, which suffices for valid and consistent subsequent inference.

\begin{propositi}(MFTDN identifiability)\label{prop:identifiability}
Suppose an MFTDN can be represented by two sets of parameters, $(\mx, \my, \mr^{[t]}_s)$ and $(\mx', \my', (\mr')^{[t]}_s)$, such that they yield the same probability matrices $\{\mpp^{[t]}_s\}_{t \in \mathcal{T}, s \in [K]}$. If $\mr^{[t]}_s$ satisfies
\begin{equation}\label{eq:rank R}
	\operatorname{rank} \left( \left[ \mr_1^{[t_1]}, \mr_1^{[t_2]}, \dots, \mr_K^{[t_m]} \right] \right) = d,
\end{equation}
where the matrix on the {\red left} is the horizontal concatenation of all $\{\mr_s^{[t]}\}_{s\in[K],t\in\mathcal{T}}$, then there exist $\mw_\mx, \mw_\my \in \mathcal{O}_d$ such that
$$\mx' = \mx \mw_\mx,\quad\my' 
= \my \mw_\my
\text{, and }
(\mr')_s^{[t]} 
= \mw_\mx^\top \mr_s^{[t]} \mw_\my.$$
\end{propositi}

The condition in \eqref{eq:rank R} guarantees that, even if individual matrices $\tilde\mpp^{[t]}_s = \mx \mr^{[t]}_s \my^\top$ may have rank less than $d$, the latent structure is still identifiable up to orthogonal transformations as long as $\{\mr^{[t]}_s\}$ across all time points and layers jointly achieves full rank $d$.

As orthogonal transformations are rotations in space, and the Frobenius and spectral norms are invariant under orthogonal (or orthonormal) transformations, distance-based inference procedures can be effectively applied.
In particular, the distances $\|\mathbf{x}_i - \mathbf{x}_j\|$, $\|\mathbf{y}_i - \mathbf{y}_j\|$ can be used to measure dissimilarity between vertices $i$ and $j$ in terms of their outgoing and incoming characteristics, respectively, allowing for vertex community detection or clustering using methods such as $k$-means based on the rows of $\mx$ or $\my$.
Similarly, the distance $\|\mr_s^{[t]} - \mr_{s'}^{[t']}\|_F$ can be used to quantify dissimilarity between different time points and/or network layers. This facilitates the visualization of temporal dynamics and inter-layer differences through distance-aware dimensionality reduction techniques such as multidimensional scaling (MDS) and uniform manifold approximation and projection (UMAP). 
Moreover, for each layer $s$, the temporal trajectory $\mr_s^{[t]}$ as $t$ varies can be interpreted as a curve in $\RR^{d^2}$, providing a geometric perspective on the evolution of each network layer over time and enabling comparison and evaluation of similarities among layers.
We demonstrate these inference procedures using estimations through simulations and real data analyses presented in Sections~\ref{sec:simu} and \ref{sec:real}.

%\subsection{Related works}

\section{Estimation via Functional Tensor Tucker Decomposition and RKHS}\label{sec:estimation}

In this section, we derive an algorithm to estimate $(\mx, \my, \mr^{[t]}_s)$ of the MFTDN model from the observed adjacency matrices $\{\ma^{[t]}_s\}_{s \in [K],t \in \mathcal{T}}$, based on a functional tensor Tucker decomposition and a RKHS framework.

\subsection{Likelihood formulation and RKHS-constrained estimation}

The edge probability $\tilde\mpp^{[t]}$ defined in Section \ref{sec:model} can be viewed as a functional tensor as introduced in Section \ref{sec:notation}. 
The equation $\tilde\mpp_s^{[t]} = \mx \mr_s^{[t]} \my^\top$ in Definition \ref{def_multilayer_ftdn} is equivalent to say the functional tensor $\tilde\mpp \in \RR^{K \times n \times n \times [0,1]}$ has Tucker rank $(K, d, d)$ and admits the Tucker decomposition
\[
    \tilde\mpp^{[t]} = \mr^{[t]} \times_2 \mx \times_3 \my.  
\]
A natural estimator of $(\mx, \my, \mr)$ is MLE. Specifically, the negative log-likelihood can be computed as 
\beas
F(\mx, \my, \mr)
&= & - \sum_{s \in [K]} \sum_{t \in \mathcal{T}} \sum_{i,j\in[n]} \left[ \ma^{[t]}_{s,i,j} \log \mpp^{[t]}_{s,i,j} + (1 - \ma^{[t]}_{s,i,j}) \log(1 - \mpp^{[t]}_{s,i,j}) \right]\\
&= & - \sum_{s \in [K]} \sum_{t \in \mathcal{T}} \sum_{i,j\in[n]} \left[ \ma^{[t]}_{s,i,j} \tilde\mpp^{[t]}_{s,i,j} - \log\( 1 + \exp(\tilde\mpp^{[t]}_{s,i,j})\) \right] \\
&= & \sum_{s \in [K]} \sum_{t \in \mathcal{T}} \sum_{i,j\in[n]} f(\ma^{[t]}_{s,i,j}, \tilde\mpp^{[t]}_{s,i,j}) \\
&= & \sum_{s \in [K]} \sum_{t \in \mathcal{T}} \sum_{i,j\in[n]} f\(\ma^{[t]}_{s,i,j}, \sum_{k \in [d]} \sum_{\ell \in [d]}(\mathbf{x}_i)_k \, \mr^{[t]}_{s, k,\ell} \, (\mathbf{y}_j)_\ell\),
\eeas
where $f(x,y):= \log(1+\exp(y)) - xy$.
Hence, the MLE is the solution to 
\bea\label{eq_MLE}
     \argmin_{\mx, \my, \mr} \sum_{s \in [K]} \sum_{t \in \mathcal{T}} \sum_{i,j\in[n]} f\(\ma^{[t]}_{s,i,j}, \sum_{k \in [d]} \sum_{\ell \in [d]}(\mathbf{x}_i)_k \, \mr^{[t]}_{s, k,\ell} \, (\mathbf{y}_j)_\ell\). 
\eea
However, the optimization problem in \eqref{eq_MLE} is infinite-dimensional due to the functional nature of the parameter $\mr^{[t]}_{s,k,\ell}$. To address this, we adopt an RKHS framework for $\mr$. Specifically, we assume that each entry function $\mr_{s,k,\ell}$ lies in an RKHS $\mathcal{H}$ with reproducing kernel $\mathbb{K}(\cdot, \cdot)$. To promote smoothness and prevent overfitting, we impose the constraint
\[
    \sum_{s \in [K],\; k,\ell \in [d]}  \|\mx_{:k}\|^2\, \|\my_{:\ell}\|^2\,\|\mr_{s,k,\ell}\|_{\mathcal{H}}^2 \leq C,
\]
where $\mx_{:k}$ and $\my_{:\ell}$ denote the $k$th and $\ell$th columns of $\mx$ and $\my$, respectively, and $\|\cdot\|_{\mathcal{H}}$ is the RKHS norm.
Under this framework, the MLE problem in \eqref{eq_MLE} can be reformulated as
\bea\label{eq_optimization_problem}
     \argmin_{{\substack{\mx, \my \in \RR^{n \times d}, \mr_{s, k,\ell} \in \mathcal{H} \\ \sum_{\ell,k\in[d], s\in[K]} \|\my_{:\ell}\|^2\|\mx_{:k}\|^2\|\mr_{s, k,\ell}\|_{\mathcal{H}}^2 \leq C}}} 
     \sum_{s \in [K]}\sum_{t \in \mathcal{T}} \sum_{i,j\in[n]} f\(\ma^{[t]}_{s,i,j}, \sum_{\ell,k\in[d]} (\mathbf{x}_i)_k \, \mr^{[t]}_{s, k,\ell} \, (\mathbf{y}_j)_\ell\). 
\eea

A generalized form of the Representer Theorem \citep{scholkopf2001generalized} provides a fundamental representation of the solution to \eqref{eq_optimization_problem}, listed in the following proposition.

\begin{propositi} \label{proposition_representer_thm}
    If $f(x,y)$ is a convex and Gateaux-differentiable function of $y$ for given $x$, then the optimization problem \eqref{eq_optimization_problem} admits a solution in which $\hat \mr^{[t]}_{s, k,\ell}$ can be represented as 
    \bea\label{eq_R_representation}
        \hat \mr^{[t]}_{s, k,\ell} = \sum_{h \in \mathcal{T}} \theta_{s, k,\ell, h} \mathbb{K}(t, h)
    \eea
    for some $\theta=(\theta_{s, k,\ell, h})\in \RR^{d^2Km}$. 
\end{propositi}

With Proposition \ref{proposition_representer_thm}, we can now rewrite \eqref{eq_optimization_problem} as a discretized optimization problem: 
\bea\label{eq_optimization_problem_2}
     \argmin_{{\substack{\mx, \my \in \RR^{n \times d}, \theta\in \RR^{d^2Km} \\ \sum_{s \in [K]}\sum_{\ell,k\in[d]} \|\my_{:\ell}\|^2\|\mx_{:k}\|^2\|\mr_{s,k,\ell}\|_{\mathcal{H}}^2 \leq C}}} 
     \sum_{s \in [K]}\sum_{t \in \mathcal{T}} \sum_{i,j\in[n]} f\(\ma^{[t]}_{s,i,j}, \sum_{\ell,k\in[d]} (\mathbf{x}_i)_k (\mathbf{y}_j)_\ell \sum_{h \in \mathcal{T}} \theta_{s, k,\ell, h} \mathbb{K}(t, h) \). 
\eea
We additionally have $\|\mr_{s,k,\ell}\|_{\mathcal{H}}^2 = \|\sum_{h \in \mathcal{T}} \theta_{s, k,\ell, h} \mathbb{K}(\cdot, h)\|_\mathcal{H}^2 = \sum_{h,q \in \mathcal{T}} \theta_{s, k,\ell, h} \theta_{s, k,\ell, q} \mathbb{K}(q, h)$ by the property of RKHS. 

\subsection{Optimization algorithm and implementation}\label{sec:opti}

{\red
Now, estimating the parameters in the MFTDN reduces to the finite-dimensional optimization problem \eqref{eq_optimization_problem_2}. Consequently, one may apply any suitable optimization algorithm to solve it. In this section, we focus on the most basic approach--projected gradient descent, where the projection enforces the required constraints. 
}

Note that we have
\[
    \tilde\mpp^{[t]}_{s,i,j}=\sum_{k \in [d]} \sum_{\ell \in [d]}(\mathbf{x}_i)_k \, \mr^{[t]}_{s,k,\ell} \, (\mathbf{y}_j)_\ell = \sum_{k \in [d]} \sum_{\ell \in [d]}(\mathbf{x}_i)_k (\mathbf{y}_j)_\ell \sum_{h \in T} \theta_{s,k,\ell,h} \mathbb{K}(t, h);
\]
\[
    \frac{\partial\tilde\mpp^{[t]}_{s,i,j}}{\partial \mx_{i,k}} = \sum_{\ell \in [d]} \mr^{[t]}_{s,k,\ell} \my_{j,\ell}, \quad
    \frac{\partial\tilde\mpp^{[t]}_{s,i,j}}{\partial \my_{j,\ell}} = \sum_{k \in [d]} \mr^{[t]}_{s,k,\ell} \mx_{i,k}, \quad \text{and} \quad
    \frac{\partial\tilde\mpp^{[t]}_{s,i,j}}{\partial \theta_{s,k,\ell,h}} = \mathbb{K}(t, h) \my_{j,\ell}\mx_{i,k};
\]
and
\[\red
    \frac{\partial F}{\partial \mx} = \frac{\partial F}{\partial \tilde\mpp}\frac{\partial \tilde\mpp}{\partial \mx}, \quad
    \frac{\partial F}{\partial \my} = \frac{\partial F}{\partial \tilde\mpp}\frac{\partial \tilde\mpp}{\partial \my}, \quad
    \frac{\partial F}{\partial \theta} = \frac{\partial F}{\partial \tilde\mpp}\frac{\partial \tilde\mpp}{\partial \theta}, \quad \text{and} \quad
    \frac{\partial F}{\tilde{\mathbf P}^{[t]}_{s,i,j}} = \sigma(\tilde \mpp_{s,i,j}^{[t]})-\ma_{s,i,j}^{[t]}. 
\]
where $\sigma(x) = \exp(x) / (1 + \exp(x))$ is the logistic function. 
Hence, we can write down the gradient as
\bea\label{eq_gradient}
\begin{split}
    &\frac{\partial F}{\partial \mx_{i,k}} = \sum_{\substack{j \in [n], \ell \in [d]\\ t \in \mathcal{T}, s \in [K]}} \(\sigma(\tilde \mpp_{s,i,j}^{[t]})-\ma_{s,i,j}^{[t]}\) \mr^{[t]}_{s,k,\ell} \my_{j,\ell}, \\
    &\frac{\partial F}{\partial \my_{j,\ell}} = \sum_{\substack{i \in [n], k \in [d]\\ t \in \mathcal{T}, s \in [K]}}  \(\sigma(\tilde \mpp_{s,i,j}^{[t]})-\ma_{s,i,j}^{[t]}\) \mr^{[t]}_{s,k,\ell} \mx_{i,k}, \\
    &\frac{\partial F}{\partial \theta_{s,k,\ell,h}} = \sum_{\substack{i, j \in [n]\\ t \in \mathcal{T}}} \(\sigma(\tilde \mpp_{s,i,j}^{[t]})-\ma_{s,i,j}^{[t]}\) \mathbb{K}(t, h) \my_{j,\ell}\mx_{i,k}. 
\end{split}
\eea
For some learning rate $\alpha$, we update via
\[
    \hat \mx \leftarrow \hat \mx  - \alpha \frac{\partial F}{\partial \mx}, \quad
    \hat \my \leftarrow \hat \my  - \alpha \frac{\partial F}{\partial \my},  \quad \text{and} \quad
    \hat \theta \leftarrow \hat \theta  - \alpha \frac{\partial F}{\partial \theta}. 
\]
To ensure $\sigma^2 := \sum_{s \in [K]}\sum_{k,\ell\in[d]} \|\mx_{:k}\|^2\|\my_{:\ell}\|^2\|\mr_{s,k,\ell}\|_{\mathcal{H}}^2 \leq C^2$, we let 
\[
    \hat \mx \leftarrow C^{1/3} \hat \mx / \sigma^{1/3} , \quad
    \hat \my \leftarrow C^{1/3} \hat \my / \sigma^{1/3} , \quad \text{and} \quad
    \hat \theta \leftarrow C^{1/3} \hat \theta / \sigma^{1/3} ,
\]
if $\sigma > C$ after the gradient update at each iteration. 
Finally, after finishing the gradient descent, to output the $\hat \mx, \hat \my \in \mathcal{O}_{n\times d}$, we orthogonalize $\hat \mx, \hat \my$ and adjust $\hat \mr$ via 
\bea\label{eq_orthogonalize}
    \hat \mx \leftarrow \operatorname{SVD}_{d}(\hat \mx), \quad
    \hat \my \leftarrow \operatorname{SVD}_{d}(\hat \my), \quad \text{and} \quad
    \hat \mr^{[t]}_s \leftarrow \operatorname{SVD}_{d}(\hat \mx)^\T \hat \mx \hat \mr^{[t]}_s \hat \my^\T  \operatorname{SVD}_{d}(\hat \my). 
\eea
We summarize the procedure in Algorithm~\ref{algorithm_rkhs}. 

\begin{algorithm}[h]
\caption{MFTDN estimation via RKHS}
\label{algorithm_rkhs}
\KwIn{Adjacency matrices $\{\ma^{[t]}_s\}_{s\in[K],t\in\mathcal{T}}$, embedding dimension $d$, kernel $\mathbb{K}$, learning rate $\alpha$, constraint $C$, initialization $\hat\mx, \hat\my, \hat\mr$}

\For{$l = 0, 1, 2, \dots$}{
        Calculate $\frac{\partial F}{\partial \mx}$, $\frac{\partial F}{\partial \my}$, and $\frac{\partial F}{\partial \theta}$ by \eqref{eq_gradient}; 
        
        Let $\hat \mx \leftarrow \hat \mx  - \alpha \frac{\partial F}{\partial \mx}, \quad
    \hat \my \leftarrow \hat \my  - \alpha \frac{\partial F}{\partial \my},  \quad \text{and} \quad
    \hat \theta \leftarrow \hat \theta  - \alpha \frac{\partial F}{\partial \theta}$;

        Calculate $\hat \mr^{[t]}_{s,k,\ell} = \sum_{h \in \mathcal{T}} \theta_{s, k,\ell, h} \mathbb{K}(t, h)$;
        
        \If{$\sigma^2 := \sum_{s \in [K]}\sum_{k,\ell\in[d]} \|\hat\mx_{:k}\|^2\|\hat\my_{:\ell}\|^2\|\hat\mr_{s,k,\ell}\|_{\mathcal{H}}^2 > C^2$} {$\hat \mx \leftarrow C^{1/3} \hat \mx / \sigma^{1/3} , \quad
    \hat \my \leftarrow C^{1/3} \hat \my / \sigma^{1/3} , \quad \text{and} \quad
    \hat \theta \leftarrow C^{1/3} \hat \theta / \sigma^{1/3}$}
}
Orthogonalize $\hat\mx, \hat\my$ and adjust $\hat\mr$ by \eqref{eq_orthogonalize}. 

\KwOut{$\hat\mx, \hat\my$, and $\hat\mr$}
\end{algorithm}

The kernel $\mathbb{K}$ and embedding dimension $d$ can be selected by the Bayesian Information Criterion (BIC). More specifically, we calculate the BIC scores for a range of candidate hyperparameters and select the one that minimizes the BIC value:
\[
    \text{BIC} = 2 F + d(2n+dKm) \cdot \log(Kmn^2).
\]
Here, $F$ is the loss function in the optimization problem \eqref{eq_optimization_problem_2}, which depends on $\mathbb{K}$ and $d$. 
This approach balances model fit and complexity, selecting a dimension that best explains the data while avoiding overfitting. 

As discussed above and in \cite{tang2024tensor}, the constraint parameter $C$ can encourage smoothness and avoid overfitting of the functionals. Additionally, prior knowledge can also be incorporated for more efficient parameter estimation. For example, if we know the period of the MFTDN, we can choose the periodic kernel with the corresponding period. 

\begin{rema}\label{rmk:symmetric}
The MFTDN model and the estimation method in Algorithm \ref{algorithm_rkhs} can be adapted to undirected networks by imposing $\mx = \my$, requiring each $\mr^{[t]}_s$ to be symmetric, and assuming that the observed adjacency matrices $\ma^{[t]}_s$ are symmetric with independent entries in the upper triangular part. The estimation Algorithm~\ref{algorithm_rkhs} can be correspondingly modified by only updating $\mx$ and $\mr^{[t]}_{s,k,l}$ for $k \leq l$ and then assign $\my = \mx$ and $\mr^{[t]}_{s,k,l} = \mr^{[t]}_{s,l,k}$ for $k > l$.  
\end{rema}
{\color{black}
\begin{rema}\label{remark_normalization}
    Algorithm \ref{algorithm_rkhs} enforces constraint $C$ by normalizing the products of $\mx$, $\my$, and $\mr$ instead of normalizing them separately. This follows the same idea as in \cite{tang2024tensor}: empirically, we observed that separate normalization can lead to instability--it often only rescales the dominant factor, allowing the smaller factor to remain small, or even shrink further over iterations. 
    The normalization step in Algorithm~\ref{algorithm_rkhs} should be viewed as a practical stabilization heuristic. A rigorous projected-gradient guarantee can be obtained if this step is replaced by the exact Euclidean projection onto a convex constraint set. In our numerical experiments we take $C=\infty$, so the algorithm reduces to standard gradient descent on the finite-dimensional objective in $(\mx,\my,\theta)$. We provide convergence guarantees for both cases in Propositions~\ref{proposition_convergence} and \ref{proposition_convergence_2} in Appendix~\ref{sec:convergence}. 
\end{rema}
\begin{rema}\label{remark_choice_C}
The constraint $C$ in Algorithm \ref{algorithm_rkhs} plays an important role in connecting \eqref{eq_MLE} with \eqref{eq_optimization_problem_2}. In particular, for any choice of $C$, the Karush--Kuhn--Tucker conditions together with the representer theorem imply that the solution lies in the finite-dimensional subspace of the RKHS, as in Proposition \ref{proposition_representer_thm}.

The regularization effect of this RKHS formulation should be interpreted with some care. Let $G=(\mathbb K(t_a,t_b))_{a,b=1}^m$ be the kernel Gram matrix on the observed time points. For fixed $\mx$ and $\my$, the fitted values of each trajectory satisfy
\[
    (\mr_{s,k,\ell}^{[t_1]},\ldots,\mr_{s,k,\ell}^{[t_m]})^\top
    =
    G(\theta_{s,k,\ell,t_1},\ldots,\theta_{s,k,\ell,t_m})^\top .
\]
Hence, if $G$ is nonsingular and $C=\infty$, then the RKHS representation can realize arbitrary values at the observed time points. In this representation-level sense, optimizing over $\theta$ is equivalent to optimizing over unrestricted matrices $\{\mathbf R_s^{[t_a]}\}_{a=1}^m$ at the design points, so the RKHS representation provides smooth interpolation rather than local averaging at those points. However, this equivalence does not imply that the implemented MFTDN estimator coincides with a no-smoothing baseline in practice. See Appendix~\ref{app:obs_R_accuracy} for more details. 

Thus, the role of $C$ is primarily to control smoothing at the observed time points, where smoothing means local averaging or borrowing information across nearby time points when the constraint is active. In practice, we set $C=\infty$ in both the simulation studies and the real data analysis. Empirically, the fitted functions remain smooth and achieve high predictive accuracy on unobserved time points (see Section~\ref{sec:simu_error}). Nevertheless, if local averaging across nearby time points is desired in other applications, it can be encouraged by choosing a finite value of $C$. The smoothness of the estimator also depends on the choice of kernel and its associated parameters.

%The constraint $C$ in Algorithm \ref{algorithm_rkhs} plays an important role in connecting the MLE formulation \eqref{eq_MLE} with the finite-dimensional RKHS optimization problem \eqref{eq_optimization_problem_2}. In particular, for any choice of $C$, the Karush--Kuhn--Tucker conditions together with the representer theorem imply that the solution lies in a finite-dimensional subspace of the RKHS, leading to formulation \eqref{eq_optimization_problem_2}, as in Proposition \ref{proposition_representer_thm}. 

%Once the optimization problem is restricted to this finite-dimensional subspace, the estimator already enjoys an implicit regularization effect, which prevents overfitting and unsmooth solutions even when the constraint is not explicitly enforced in \eqref{eq_optimization_problem_2}. Therefore, the role of $C$ is mainly theoretical, ensuring the equivalence between the constrained MLE formulation and the finite-dimensional RKHS representation.

%In practice, we set $C=\infty$ in both the simulation studies and the real data analysis. Empirically, the resulting fitted functions remain smooth and achieve high predictive accuracy on time points not observed in the training data (see Section~\ref{sec:simu_error}), suggesting that no unsmoothness issue arises when $C=\infty$. Nevertheless, if unsmooth fitted functions are observed in other applications, the smoothness can still be controlled by choosing a finite value of $C$. We also note that the smoothness of the estimator depends on the choice of kernel and its associated parameters.
\end{rema}
}
\begin{rema}
    In each iteration of Algorithm \ref{algorithm_rkhs}, calculating the gradient with respect to $\mx$ requires calculating $\frac{\partial F}{\partial \mx_{i,k}}$ for $i\in[n]$ and $k\in[d]$ and each $\frac{\partial F}{\partial \mx_{i,k}}$ requires $O(n d m K)$ floating-point operations (flops). Thus, the calculation of gradient with respect to $\mx$ costs $O(n^2 d^2 m K)$ flops. Similarly, the calculation of gradient with respect to $\my$ and $\theta$ costs $O(n^2 d^2 m K)$ and $O(n^2 d^2 m^2 K)$ flops, respectively. 
    Calculating the values of $\hat \mr^{[t]}_{s,k,\ell}$ requires $O(d^2 m^2 K)$ flops. 
    The calculation of $\|\mr_{s,k,\ell}\|_{\mathcal{H}}$ requires $O(m^2)$ flops, so the calculation of $\sigma$ costs $O(d^2m^2K + dn)$. 
    Hence, each iteration of Algorithm \ref{algorithm_rkhs} costs $O(n^2 d^2 m^2 K)$ flops, which is dominated by the calculation of the gradient of $\theta$. 
\end{rema}

Notably, our approach for solving the MLE problem in \eqref{eq_MLE} can be extended to accommodate more general loss functions $f$ beyond those arising from the Bernoulli distribution, as well as more general functional tensors of the form $\mpp^{[t]} = \mr^{[t]} \times_{k \in [D]} \mx_k%\in \RR^{p_1 \times \cdots \times p_D \times [0,1]}
$
%$\mathbf{T}^{[t]} = \ms^{[t]} \times_{k \in [d]} \ma_k$ 
with low Tucker rank.
This generalization is formally discussed in Section~\ref{sec:gtucker}.

%{\color{black}(mention our contribution for functional tensor Tucker decomposition estimation and compare it with existing work?) I briefly mentioned in Section~\ref{sec:gtucker}. I feel it is not the focus of our work?}

{\color{black}
\begin{rema}
\label{rem:asym}
While due to the involvement of functional tensor decomposition, establishing rigorous asymptotic analysis for MFTDN is challenging, an approximate upper error bound for MFTDN can be obtained by analyzing a baseline model without the smoothness assumption.
See Appendix~\ref{sec:asym} for detailed discussion.
\end{rema}

}

\subsection{Initialization via spectral methods}\label{sec:initial}

To accelerate the algorithm and ensure reliable estimation, we propose an initialization strategy based on spectral methods. 

\textbf{Step~1:} For each $s \in [K]$ and $t \in \mathcal{T}$, we compute $\hat{\mx}^{[t]}_s$ and $\hat{\my}^{[t]}_s$ as $n \times d$ matrices, whose columns are the top $d$ left and right singular vectors of $\ma^{[t]}_s - \frac{1}{2} \mathbf{1}\mathbf{1}^\top$, respectively.  
Then we construct $\hat{\mx}$ and $\hat{\my}$ as $n \times d$ matrices, whose columns are the $d$ leading left singular vectors of the concatenated matrices $\big[\hat{\mx}^{[t_1]}_1 \mid \hat{\mx}^{[t_2]}_1 \mid \dots \mid \hat{\mx}^{[t_m]}_K \big]$ and $\big[\hat{\my}^{[t_1]}_1 \mid \hat{\my}^{[t_2]}_1 \mid \dots \mid \hat{\my}^{[t_m]}_K \big]$, respectively.

This Step~1 computes the initial estimates $\hat\mx$ and $\hat\my$ using a variation of the COSIE model \citep{arroyo2021inference}, which does not account for temporal smoothness but provides a sufficient initialization for our Algorithm~\ref{algorithm_rkhs}. Similar ideas are widely used in subspaces capturing for multiple matrices \citep{tang2025mode, yang2025estimating}. 
Note that the COSIE model assumes $\mpp^{[t]}_s = \mx \mr^{[t]}_s \my$, so $\hat{\mx}^{[t]}_s$ and $\hat{\my}^{[t]}_s$ are computed as the leading subspaces of $\ma^{[t]}_s$ to approximate those of $\mpp^{[t]}_s$, which are then aggregated to obtain $\mx$ and $\my$.
However, in our model, we aim to use $\hat{\mx}^{[t]}_s$ and $\hat{\my}^{[t]}_s$ to estimate the leading subspaces of $\tilde{\mpp}^{[t]}_s$, rather than $\mpp^{[t]}_s$. 
Recall that $\mpp^{[t]}_s$ and $\tilde\mpp^{[t]}_s$ are related through a logit transformation, i.e., $\text{logit}(\mpp^{[t]}_s) = \tilde\mpp^{[t]}_s$. 
Fortunately, under mild conditions, such as when the entries of $\tilde\mpp^{[t]}_s$ are neither too large nor too small, the leading subspaces of $\tilde\mpp^{[t]}_s$ can be approximated from $\mpp^{[t]}_s$. %Specifically, we assume the singular value decomposition of the rank-$d$ matrix $\tilde\mpp^{[t]}_s$ is given by $\mx^{[t]}_s \mSigma^{[t]}_s \my^{[t]\top}_s$. 
Specifically, notice that the inverse logit function can be approximated linearly as $\text{logit}^{-1}(\tilde{p}) \approx \frac{1}{2} + \frac{1}{4} \tilde{p}$, when $\tilde{p}$ is not too large or small. In this case, we have
$
\mathbb{E} (\ma^{[t]}_s) = \mpp^{[t]}_s \approx \frac{1}{2} \mathbf{1}\mathbf{1}^\top + \frac{1}{4} \tilde\mpp^{[t]}_s. 
$
Thus, we subtract the term $\frac{1}{2} \mathbf{1}\mathbf{1}^\top$ from $\ma^{[t]}_s$ first, and then perform SVD to estimate the subspaces of each $\tilde{\mpp}^{[t]}_s$.
%where the additive term $\frac{1}{2} \mathbf{1}\mathbf{1}^\top$ does not affect the subspaces spanned by $\frac{1}{4} \tilde\mpp^{[t]}_s$, because $\mx^{[t]}_s$ and $\my^{[t]}_s$ are orthogonal to $\mathbf{1}$. 
%Thus when the singular values of $\frac{1}{4} \tilde\mpp^{[t]}_s$ are sufficiently large compared to that of $\frac{1}{2} \mathbf{1}\mathbf{1}^\top$, the leading subspaces of $\mpp^{[t]}_s$ and $\tilde\mpp^{[t]}_s$ are approximately identical.
%In more complex cases, for instance, when the singular values of $\frac{1}{4} \tilde\mpp^{[t]}_s$ are not sufficiently large, we can compute the $(d+1)$ leading singular subspaces of $\ma^{[t]}_s$ and then remove the $1$-dimensional subspace associated with $\mathbf{1}$ by projecting onto its orthogonal complement to get $\hat\mx^{[t]}_s$ and $\hat\my^{[t]}_s$. 
%We emphasize that this step serves as an algorithmic initialization, and an approximate estimation of the leading subspaces is sufficient for further refinement in the main algorithm. {\color{black} (please check whether you agree)}

\textbf{Step~2:} With reliable estimates of $\mx$ and $\my$ in hand, the original problem \eqref{eq_optimization_problem_2} reduces to a convex optimization problem:
\bea\label{eq_optimization_problem_R}
     \argmin_{\theta} 
     \sum_{s \in [K]}\sum_{t \in \mathcal{T}} \sum_{i,j\in[n]} f\(\ma^{[t]}_{s,i,j}, \sum_{\ell,k\in[d]} (\mathbf{x}_i)_k (\mathbf{y}_j)_\ell \sum_{h \in \mathcal{T}} \theta_{s, k,\ell, h} \mathbb{K}(t, h) \), 
\eea
which allows for efficient estimation of $\mr$. We summarize our full initialization procedure as in Algorithm \ref{algorithm_asc}. 

\begin{algorithm}[h]
\caption{Initialization}
\label{algorithm_asc}
\KwIn{Adjacency matrices $\{\ma^{[t]}_s\}_{s\in[K],t\in\mathcal{T}}$, embedding dimension $d$}

Calculate $\hat{\mx}^{[t]}_s = \SVD_d\left(\ma^{[t]}_s- \frac{1}{2} \mathbf{1}\mathbf{1}^\top\right)$ and $\hat{\my}^{[t]}_s = \SVD_d\left(\ma^{[t]\top}_s- \frac{1}{2} \mathbf{1}\mathbf{1}^\top\right)$ for $s \in [K]$ and $t \in \mathcal{T}$;

Calculate $\hat{\mx} = \SVD_d\left(%\frac{1}{K|\mathcal{T}|}\sum_{s\in[K],t\in\mathcal{T}} \hat{\mx}^{[t]}_s (\hat{\mx}^{[t]}_s)^\T\)
\big[\hat{\mx}^{[t_1]}_1 \mid \hat{\mx}^{[t_2]}_1 \mid \dots \mid \hat{\mx}^{[t_m]}_K \big]\right)$ and $\hat{\my} = \SVD_d\left(%\frac{1}{K|\mathcal{T}|} \sum_{s\in[K],t\in\mathcal{T}} \hat{\my}^{[t]}_s (\hat{\my}^{[t]}_s)^\T\)
\big[\hat{\my}^{[t_1]}_1 \mid \hat{\my}^{[t_2]}_1 \mid \dots \mid \hat{\my}^{[t_m]}_K \big]\right)
$;

Solve the convex optimization problem \eqref{eq_optimization_problem_R} and calculate $\hat \mr$ by \eqref{eq_R_representation};

\KwOut{$\hat\mx, \hat\my$, and $\hat\mr$}
\end{algorithm}

\section{Simulation Experiments}\label{sec:simu}

In this section, we study the empirical performance of Algorithm~\ref{algorithm_rkhs} in estimating the parameters of the MFTDN model, as well as its effectiveness in subsequent inference tasks, including dynamic network clustering and vertex community detection, in comparison with existing methods.

\subsection{Estimation accuracy}\label{sec:simu_error}

We first evaluate the accuracy of the estimator $(\hat\mx, \hat\my, \hat\mr^{[t]}_s)$ obtained by Algorithm~\ref{algorithm_rkhs} in recovering the true $(\mx, \my, \mr^{[t]}_s)$.

{\red We} simulate a dynamic multilayer SBM as described in Section~\ref{sec:multilayer}, consisting of $K$ layers over $n$ vertices, partitioned into $d$ communities, and is observed at $m$ equally spaced time points $\mathcal{T} = \{t_1, \dots, t_m\}$ within the interval $[0, 1]$.
The outgoing and incoming community assignments are generated as follows. For each vertex $i$, we randomly assign its outgoing community $c(i) \in [d]$, where the assignment probabilities for each community are $1/d$. The assignments for different vertices are independent. Based on $c(i)$, we construct the outgoing community assignment matrix $\mz_{\text{out}} \in \{0, 1\}^{n \times d}$ such that $(\mz_{\text{out}})_{i,k} = 1$ if and only if $k = c(i)$. Similarly, the incoming community for vertex $i$, $c'(i)$, is randomly generated with the same probability distribution, and the corresponding incoming community assignment matrix $\mz_{\text{in}} \in \{0, 1\}^{n \times d}$ is constructed such that $(\mz_{\text{in}})_{i,k} = 1$ if and only if $k = c'(i)$.
For the probabilities of edges within and between communities in different layers, we generate a time-varying matrix $\tilde{\mathbf{B}}^{[t]}_s \in \mathbb{R}^{d \times d}$ for each layer $s \in [K]$ as
$
	\tilde{\mathbf{B}}^{[t]}_{s,k,\ell}
	= \mu_{s,k,\ell}
	+ \delta_{s,k,\ell} \cdot \sin\left( 2\pi \cdot \frac{t}{M} + \phi_{s,k,\ell} \right),
$
where $M = 3$, and other parameters are sampled as 
$
\mu_{s,k,\ell} \sim \text{Unif}(-10,10), 
\delta_{s,k,\ell} \sim \text{Unif}(-1,1), 
\phi_{s,k,\ell} \sim \text{Unif}\left(0, \frac{2\pi}{2}\right).
$
We then compute the probability matrices $
\mathbf{P}^{[t]}_s = \mz_{\text{out}} \cdot \text{logit}^{-1}(\tilde{\mathbf{B}}^{[t]}_s) \cdot \mz_{\text{in}}^\top,
$
and independently generate the observed adjacency matrices $\{\ma^{[t]}_s\}_{t \in \mathcal{T}}$.
Then in this MFTDN, $\mx=\mz_{\text{out}}(\mz_{\text{out}}^\top\mz_{\text{out}})^{-1/2}$, $\my=\mz_{\text{in}}(\mz_{\text{in}}^\top\mz_{\text{in}})^{-1/2}$, and $\mr^{[t]}_s=(\mz_{\text{out}}^\top\mz_{\text{out}})^{1/2}\tilde\mb^{[t]}(\mz_{\text{in}}^\top\mz_{\text{in}})^{1/2}$.

%We randomly generate the orthonormal matrices $\mx, \my \in \mathcal{O}_{n \times d}$.
%For each layer $s \in [K]$, we construct a time-varying matrix $\mr^{[t]}_s \in \mathbb{R}^{d \times d}$ as
%$$\mr^{[t]}_{s,k,\ell} = \mu_{s,k,\ell} + \delta_{s,k,\ell} \cdot \sin\left(2\pi\cdot\frac{t}{M} + \phi_{s,k,\ell}\right),$$
%where $M = 3$, and other parameters are sampled as
%$\mu_{s,k,\ell} \sim \text{Unif}(...) \times n,
%\delta_{s,k,\ell} \sim \text{Unif}(...) \times n, 
%\phi_{s,k,\ell} \sim \text{Unif}\left(0, \frac{2\pi}{2}\right).$
%Note that $\mu_{s,k,\ell}$ and $\delta_{s,k,\ell}$ are multiplied by $n$ to ensure that the entries of $\mathbf{P}^{[t]}_s$ remain of the same order as $n$ changes, given that $\mathbf{X}, \mathbf{Y}$ are random matrices drawn from $\mathcal{O}_{n \times d}$, where each row norm is approximately $\sqrt{d/n}$.
%We then compute $\tilde\mpp^{[t]}_s = \mx \mr^{[t]}_s \my^\top$,
%and obtain
%$\mpp^{[t]}_s = \operatorname{logit}^{-1}(\tilde\mpp^{[t]}_s )$.
%Finally we generate the adjacency matrices $\{\ma^{[t]}_s\}_{s\in[K],t\in\mathcal{T}}$ by sampling each entry independently as $\ma^{[t]}_{s,i,j} \sim \text{Bernoulli}(\mpp^{[t]}_{s,i,j}).$

\begin{sloppypar}

After obtaining the estimates from Algorithm~\ref{algorithm_rkhs} with periodic kernel $\mathbb{K}(x, y) = \exp \left(-{\sin ^2(2\pi|x-y| / M)}\right)$, we measure the estimation accuracy. Considering that the MFTDN is unique up to orthogonal transformations as mentioned in Section~\ref{sec:identifiability}, we use the spectral norm of the difference between the projections of $\hat\mx$ and $\mx$, as well as $\hat\my$ and $\my$, to quantify the difference between the common invariant subspaces and their estimates, i.e.,
$
\text{Err}(\hat\mx) := \|\hat\mx\hat\mx^\top - \mx\mx^\top\|, \quad
\text{Err}(\hat\my) := \|\hat\my\hat\my^\top - \my\my^\top\|.
$
The distance is $0$ if and only if the true subspace and the estimated subspace differ only by an orthonormal transformation, while a distance of $1$ indicates that the estimated subspace is completely orthogonal to the true subspace.
Since $\mx$ and $\my$ have similar properties, we define a combined measure given by
$
\text{Err}(\hat\mx, \hat\my) = \frac{1}{2} \big(\text{Err}(\hat\mx) + \text{Err}(\hat\my)\big).
$
We next align the estimated subspaces with the true ones by solving the orthogonal Procrustes problems to obtain the orthogonal matrices $\mw_\mx$ and $\mw_\my$ as
$
\mw_\mx = \argmin_{\mo \in \mathcal{O}_d} \|\hat\mx \mo - \mx\|_F, %\quad
\mw_\my = \argmin_{\mo \in \mathcal{O}_d} \|\hat\my \mo - \my\|_F.
$
As mentioned in Section~\ref{sec:identifiability}, $\mr^{[t]}_s$ should also be aligned by applying the transformations $\mw_\mx^\top \hat\mr^{[t]}_s \mw_\my$.
Although we only observe the adjacency matrices at the time points in $\mathcal{T}$, both the true $\mr^{[t]}_s$ and the estimated $\hat\mr^{[t]}_s$ are functions of $t$ over the interval $[0,1]$. Therefore, when measuring the accuracy of $\hat\mr^{[t]}_s$, we densely select $m'=100$ evenly spaced time points within the total time interval $[0,1]$, denoted as $\mathcal{T}'=\{t_1',t_2',\dots,t_{m'}'\}$, and compare the estimated $\hat\mr^{[t]}_s$ with the true $\mr^{[t]}_s$ across all $K$ layers and $m'$ time points by
\[
\begin{aligned}
    \text{Acc}(\hat\mr) := & \operatorname{Corr}\left(
    \left[\mw_\mx^\top \hat\mr^{[t_1']}_1 \mw_\my \,|\,
    \mw_\mx^\top \hat\mr^{[t_2']}_1 \mw_\my \,|\,
    \dots \,|\,
    \mw_\mx^\top \hat\mr^{[t'_{m'}]}_K \mw_\my\right], \left[\mr^{[t_1']}_1 \,|\,
     \mr^{[t_2']}_1 \,|\,
    \dots \,|\,
     \mr^{[t'_{m'}]}_K\right]
    \right), \\
    = & \operatorname{Corr}\left(
    \mw_\mx^\top\left[ \hat\mr^{[t_1']}_1  \,|\,
   \hat\mr^{[t_2']}_1  \,|\,
    \dots \,|\,
\hat\mr^{[t'_{m'}]}_K \right]\mw_\my, 
\left[\mr^{[t_1']}_1 \,|\,
     \mr^{[t_2']}_1 \,|\,
    \dots \,|\,
     \mr^{[t'_{m'}]}_K\right]
    \right), 
\end{aligned}
\]
where the correlation between two matrices is defined as
$
\operatorname{Corr}(\mathbf{M}_1, \mathbf{M}_2) :=
\frac{\langle \operatorname{vec}(\mathbf{M}_1), \operatorname{vec}(\mathbf{M}_2) \rangle}
{\|\mathbf{M}_1\|_F \cdot \|\mathbf{M}_2\|_F}.
$
The correlation measure $\operatorname{Corr}(\mathbf{M}_1, \mathbf{M}_2)$ quantifies the similarity between the matrices $\mathbf{M}_1$ and $\mathbf{M}_2$, with values ranging from $-1$ to $1$, where $1$ indicates perfect positive correlation, $0$ indicates no correlation, and $-1$ indicates perfect negative correlation.

\end{sloppypar}

{\color{black}
We also compute the results of COSIE as a baseline, which treats all networks as exchangeable without exploiting temporal ordering, to demonstrate the advantage of MFTDN's temporal smoothness assumption. Specifically, we pool all $K$ layers across all $m$ observed time points, treating the resulting $K \times m$ networks as distinct layers in COSIE, and apply the algorithm in \cite{arroyo2021inference} to obtain $\hat{\mathbf{X}}$ and $\hat{\mathbf{Y}}$ for comparison. Note that MFTDN estimates $\hat{\mathbf{R}}$ as a continuous trajectory over $t$ and is evaluated at densely sampled time points for computing $\text{Acc}(\hat{\mathbf{R}})$, whereas COSIE only provides $\hat{\mathbf{R}}^{[t]}_s$ at the observed discrete time points. Therefore, we only compare $\text{Err}(\hat{\mathbf{X}}, \hat{\mathbf{Y}})$ for the COSIE baseline.

\begin{figure}[htbp]
    \centering
    \includegraphics[width=0.95\textwidth,height=4.3 cm,keepaspectratio=false]{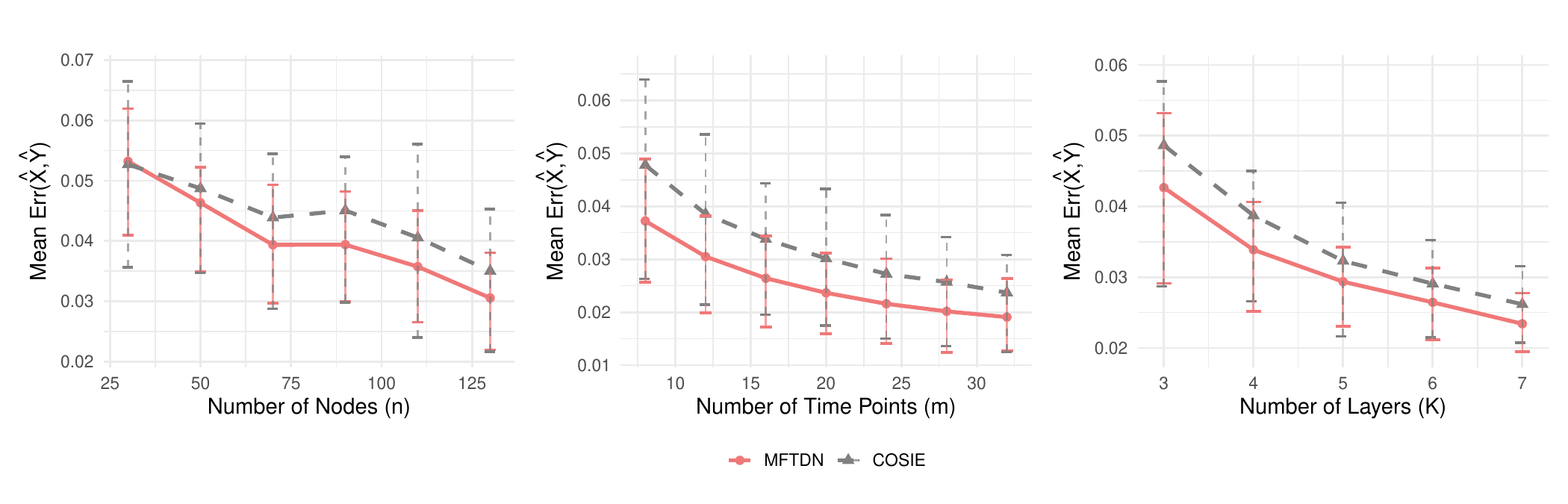}
    \includegraphics[width=0.95\textwidth,height=3.5cm,keepaspectratio=false]{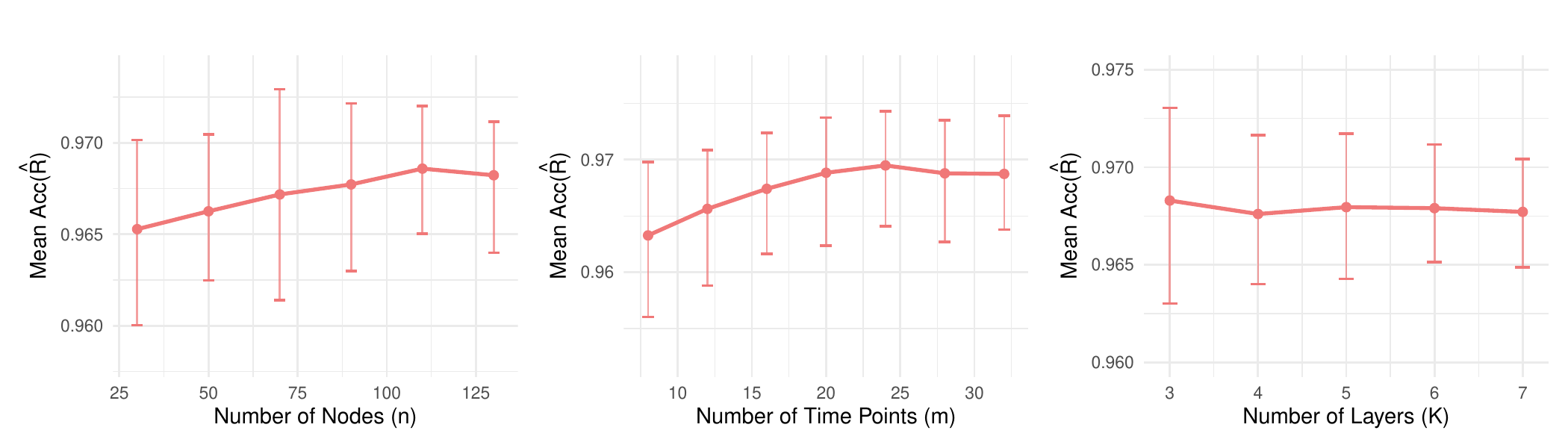}
    \caption{ Empirical $\operatorname{Err}(\hat \mx,\hat \my)$ and $\operatorname{Acc}(\hat \mr)$ for the MFTDN model estimation under the following settings: (1) varying $n \in \{30, 50, 70, 90, 110, 130\}$ while fixing $m=20$, $K=4$, and $d=3$ fixed, (2) varying $m \in \{8, 12, 16, 20, 24, 28, 32\}$ while fixing $n=100$, $K=5$, and $d=2$, and (3) varying $K \in \{3, 4, 5, 6, 7\}$ while fixing $n=100$, $m=20$, and $d=3$. Additional details of the settings are provided in Section~\ref{sec:simu_error}. The lines represent the means of $64$ independent Monte Carlo replicates, with error bars indicating the $20$th and $80$th quantiles. The total runtime for all settings is 1.5 hours using 64 parallel computations. The gray dashed lines with error bars show the results of the COSIE model for comparison.}
    \label{fig:simu_error}
\end{figure}
}

Figure~\ref{fig:simu_error} shows the results of varying the number of vertices $n$, the number of time points $m$, and the number of layers $K$.
The estimation accuracy of $\mx$ and $\my$ improves as $n$, $m$, or $K$ increases, since the subspaces are shared by all time points and layers. 
For $\mathbf{R}$, note that different layers $s$ have distinct $\mathbf{R}^{[t]}_s$, so increasing $K$ does not improve the estimation accuracy of $\mathbf{R}$, while more vertices or denser observed time points can improve the estimation.
{\color{black}Moreover, MFTDN outperforms COSIE in $\text{Err}(\hat{\mathbf{X}}, \hat{\mathbf{Y}})$, demonstrating the benefit of leveraging temporal smoothness.
}

\subsection{Dynamic network clustering}\label{sec:multi}

We consider using the estimated heterogeneous matrices $\{\hat\mr^{[t]}_s\}$ obtained from Algorithm~\ref{algorithm_rkhs} to cluster the dynamic layers across different $s \in [K]$. 

We simulate a dynamic multiple network under the MFTDN model with $K$ layers over $n$ vertices, observed at $m$ equally spaced time points $\mathcal{T}$ in $[0,1]$, where each layer $s \in [K]$ is randomly assigned to one of three clusters, indexed by $c(s) \in \{1, 2, 3\}$, and all layers within the same cluster share a common latent dynamic evolution pattern $\{\mpp^{[t]}_c\}_{t}$.
We randomly generate the orthonormal matrices $\mx, \my \in \mathcal{O}_{n \times d}$ with the embedding dimension $d = 2$. For each cluster $c \in \{1,2,3\}$, we construct a time-varying matrix $\mr^{[t]}_c \in \mathbb{R}^{d \times d}$ as
$\mr^{[t]}_{c,k,\ell} = \mu_{c,k,\ell} + \delta_{c,k,\ell} \cdot \sin\left(2\pi\cdot\frac{t}{M} + \phi_{c,k,\ell}\right),$
where $M = 3$, and other parameters are independently sampled as
$\mu_{c,k,\ell} \sim \text{Unif}(0.15, 0.15) \times n,
\delta_{c,k,\ell} \sim \text{Unif}(0.15, 0.15) \times n, 
\phi_{c,k,\ell} \sim \text{Unif}\left(0, \frac{2\pi}{2}\right).$
Then, for each $s \in [K],t \in \mathcal{T}$, we compute $\tilde\mpp^{[t]}_s=\mx \mr^{[t]}_{c(s)} \my^\top$ and obtain the probability matrix $\mpp^{[t]}_s = \operatorname{logit}^{-1}(\tilde\mpp^{[t]}_s)$. Finally, we independently generate the corresponding adjacency matrices $\{\ma^{[t]}_s\}_{s \in [K],t \in \mathcal{T}}$.

%After obtaining the estimates $\{\hat\mr^{[t]}_s\}_{s \in [K], t \in [0,1]}$ from Algorithm~\ref{algorithm_rkhs}, we treat the dynamic matrix trajectory $\{\hat\mr_s^{[t]}\}_{t \in [0,1]}$ for each layer $s$ as a curve in $\mathbb{R}^{d^2}$.
%To cluster the $K$ layers, we consider {\color{black}20 (is it correct?)} equally spaced time points in $[0,1]$ and concatenate all entries of ${\hat\mr_s^{[t]}}$ at these time points into a single feature vector for each layer $s$.
After obtaining the estimates $\{\hat{\mathbf{R}}^{[t]}_s\}_{s \in [K], t \in [0,1]}$ from Algorithm~\ref{algorithm_rkhs} with periodic kernel $\mathbb{K}(x, y) = \exp \left(-{\sin ^2(2\pi|x-y| / M)}\right)$, we note that for each layer $s$, every entry $(k,\ell)$ of $\hat{\mathbf{R}}^{[t]}_s$ is a function of $t$. According to Proposition~\ref{proposition_representer_thm}, each such function is a linear combination of the kernel $\mathbb{K}(t,h)$, with coefficients $\{\theta_{s,k,\ell,h}\}_{h \in \mathcal{T}}$.
To cluster the $K$ layers, we concatenate all coefficients $\{\theta_{s,k,\ell,h}\}_{k,\ell,h}$ into a single feature vector for each layer $s$. 
We then apply $k$-means clustering with $k=3$ to these feature vectors.
To evaluate the performance, we compute the clustering accuracy as the proportion of layers correctly assigned to their true cluster, up to a permutation of cluster labels.

%We also compute the results of COSIE for comparison. Specifically, after obtaining $\{\hat{\mr}^{[t]}_s\}_{s \in [m], t \in [T]}$ from COSIE, we cluster the $K$ layers by concatenating all entries of $\{\hat{\mr}_s^{[t]}\}_{t \in \mathcal{T}}$ into a single feature vector for each layer $s$, and then applying $k$-means to these feature vectors.

{\color{black}
We also compute the results of two baseline methods for comparison: COSIE and a version of MFTDN without temporal smoothing (baseline w/o Smoothing); more details on baseline w/o Smoothing are provided in Appendix~\ref{sec:no-smooth}. Both methods do not utilize temporal smoothing constraints, treating all observed networks for different $(s, t) \in [K]\times \mathcal{T}$ as independent layers, and the main difference is that baseline w/o Smoothing accounts for the logistic link function, while COSIE does not. After obtaining $\{\hat{\mathbf{R}}^{[t]}_s\}_{s \in [K], t \in \mathcal{T}}$ from each method, we cluster the $K$ layers by concatenating all entries of $\{\hat{\mathbf{R}}_s^{[t]}\}_{t \in \mathcal{T}}$ into a single feature vector for each layer $s$, and then applying $k$-means to these feature vectors. In our experimental setting, the linear predictors $\tilde{\mathbf{P}}^{[t]}_s$ are low-rank while the probability matrices $\mathbf{P}^{[t]}_s$ are not, making baseline w/o Smoothing yield better performance than COSIE and more appropriate as a baseline.

\begin{figure}[htbp]
    \centering
    \includegraphics[width=0.99\textwidth]{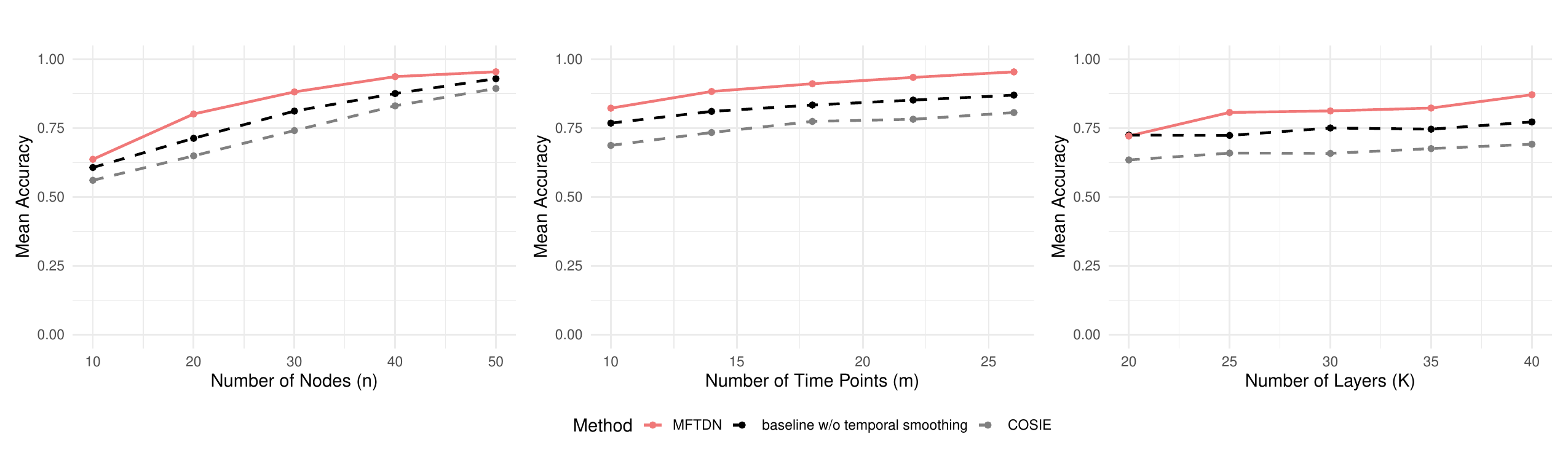}
    \caption{Empirical accuracy, measured as the proportion of layers clustered correctly, using the MFTDN model estimation under the following settings: (1) varying $n \in \{10, 20, 30, 40, 50\}$ while fixing $m=15$ and $K=10$, (2) varying $m \in \{10, 14, 18, 22, 26\}$ while fixing $n=30$ and $K=10$, and (3) varying $K \in \{20, 25, 30, 35, 40\}$ while fixing $n=20$ and $m=10$. Additional details of the settings are provided in Section~\ref{sec:multi}. 
    The results are means of $256$ independent Monte Carlo replicates.
    %The red lines represent the means of $256$ independent Monte Carlo replicates, with error bars indicating the $20$th and $80$th quantiles. 
    The gray and black dashed lines %with error bars
     show the results of COSIE and MFTDN without temporal smoothing, respectively, for comparison.}
    \label{fig:simu_multi}
\end{figure}
}

%Figure~\ref{fig:simu_multi} %shows the results of varying the number of vertices $n$, the number of time points $m$, and the number of layers $K$, demonstrating 
%demonstrates that accuracy increases as $n$, $m$, or $K$ grows. 
%Recall that COSIE ignores the temporal ordering of networks observed at different time points, while MFTDN leverages the smoothness of dynamic networks over time. As shown in Figure~\ref{fig:simu_multi}, MFTDN achieves more accurate layer clustering results.

Figure~\ref{fig:simu_multi} demonstrates that accuracy increases as $n$, $m$, or $K$ grows, and MFTDN achieves the most accurate layer clustering results. Comparing MFTDN to baseline w/o Smoothing demonstrates the substantial improvement gained by exploiting temporal smoothness, while the comparison with COSIE shows the combined benefit of both the correct link function and temporal smoothness.

\subsection{Vertex community detection}\label{sec:dynamic SBM}

We consider the dynamic SBM, as mentioned in Section~\ref{sec:single}, to evaluate the performance of FTDN in vertex community detection based on $\hat{\mathbf{X}}$ and $\hat{\mathbf{Y}}$ obtained from Algorithm~\ref{algorithm_rkhs}, and compare it with other existing methods capable of vertex community detection for dynamic networks.

We simulate a dynamic SBM on $n$ vertices with $d = 3$ communities, observed at $m$ equally spaced time points $\mathcal{T}$ in $[0, 1]$. 
We randomly generate the outgoing and incoming community assignment matrices, $\mz_{\text{out}}$ and $\mz_{\text{in}}$, as described in Section~\ref{sec:simu_error}.
%The outgoing and incoming community assignments are generated as follows. For each vertex $i$, we randomly assign its outgoing community $c(i) \in [d]$, where the assignment probabilities for each community are $1/d$. The assignments for different vertices are independent. Based on $c(i)$, we construct the outgoing community assignment matrix $\mz_{\text{out}} \in \{0, 1\}^{n \times d}$ such that $(\mz_{\text{out}})_{i,k} = 1$ if and only if $k = c(i)$. Similarly, the incoming community for vertex $i$, $c'(i)$, is randomly generated with the same probability distribution, and the corresponding incoming community assignment matrix $\mz_{\text{in}} \in \{0, 1\}^{n \times d}$ is constructed such that $(\mz_{\text{in}})_{i,k} = 1$ if and only if $k = c'(i)$.
For the probabilities of edges within and between the communities, we generate a time-varying matrix $\tilde{\mathbf{B}}^{[t]} \in \mathbb{R}^{d \times d}$ as
$
	\tilde{\mathbf{B}}^{[t]}_{k,\ell}
	= \mu_{k,\ell}
	+ \delta_{k,\ell} \cdot \sin\left( 2\pi \cdot \frac{t}{M} + \phi_{k,\ell} \right),
$
where $M = 3$, and other parameters are sampled as 
$
\mu_{k,\ell} \sim \text{Unif}(-1,1), 
\delta_{k,\ell} \sim \text{Unif}(-1,1), 
\phi_{k,\ell} \sim \text{Unif}\left(0, \frac{2\pi}{4}\right).
$
We then compute  $
\mathbf{P}^{[t]} = \mz_{\text{out}} \cdot \text{logit}^{-1}(\tilde{\mathbf{B}}^{[t]}) \cdot \mz_{\text{in}}^\top,
$
and independently generate the observed adjacency matrices $\{\ma^{[t]}\}_{t \in \mathcal{T}}$.

\begin{sloppypar}
After obtaining $\hat{\mx}$ and $\hat{\my}$ from Algorithm~\ref{algorithm_rkhs} with periodic kernel $\mathbb{K}(x, y) = \exp \left(-{\sin ^2(2\pi|x-y| / M)}\right)$, we use the rows of $\hat{\mx}$ to perform $k$-means clustering with $k = 3$ to obtain the estimated outgoing community assignments. The clustering accuracy, $\text{accuracy}_\text{out}$, is then evaluated as the proportion of vertices correctly assigned to their true communities, up to a permutation of community labels. Similarly, using the rows of $\hat{\my}$, we obtain the estimated incoming community assignments and compute the corresponding accuracy, $\text{accuracy}_\text{in}$.
We measure the overall clustering accuracy as the average of the outgoing and incoming clustering accuracies, i.e.
$
\text{overall accuracy} = \frac{1}{2} \left( \text{accuracy}_{\text{out}} + \text{accuracy}_{\text{in}} \right).
$

We compare our method with the following existing methods.
(1) Initialization of Algorithm~\ref{algorithm_rkhs}. Using the initialized $\hat\mx$ and $\hat\my$ obtained from Section~\ref{sec:initial}, $k$-means clustering is performed on their rows to obtain clustering results.
(2) COSIE, where $\mpp^{[t]} = \mx\mr^{[t]}\my$ for $t \in \mathcal{T}$. After obtaining $\hat{\mx}$ and $\hat{\my}$ using spectral methods, $k$-means clustering is performed on their rows to produce the clustering results.  
(3) The latent process model for functional networks \citep{macdonald2022latent,shi2025exploratory}, where $\mpp^{[t]} = \mx^{[t]} \mx^{[t]\top}$. Here, $\mx^{[t]} \in \mathbb{R}^{n \times d}$ represents smooth latent positions of vertices over time $t$, modeled as functional adjacency spectral embedding (FASE). The estimated FASE, $\hat{\mx}^{[t]}$, is obtained using B-spline methods.  
Note that the original model is designed for undirected networks, where $\mpp^{[t]}$ is symmetric. We extend the model to directed networks with $\mpp^{[t]}=\mx^{[t]} \my^{[t]\top}$. %$k$-means clustering is then performed on the rows of $\hat{\mx}^{[t]}$ and $\hat{\my}^{[t]}$ to produce the clustering results.
{\color{black}
To perform clustering, we concatenate the estimated latent positions across all observed time points for each vertex, and apply $k$-means clustering to the concatenated feature vectors $\{[\hat{\mathbf{x}}_{i}^{[t_1]} \mid \cdots \mid \hat{\mathbf{x}}_{i}^{[t_m]}]\}_{i \in [n]}$ and $\{[\hat{\mathbf{y}}_{i}^{[t_1]} \mid \cdots \mid \hat{\mathbf{y}}_{i}^{[t_m]}]\}_{i \in [n]}$.
}
(4) The latent space model (LSM) for dynamic networks \citep{sewell2015latent}, where  
$
\tilde{\mpp}_{ij}^{[t]} = \beta_{\text{IN}} \left(1 - \frac{d_{ij}^{[t]}}{r_j}\right) 
+ \beta_{\text{OUT}} \left(1 - \frac{d_{ij}^{[t]}}{r_i}\right),
$
with $d_{ij}^{[t]} = \|\mathbf{x}_{i}^{[t]} - \mathbf{x}_{j}^{[t]}\|$, and  
$
\mathbf{x}_{i}^{[t_\ell]} \mid \mathbf{x}_{i}^{[t_{\ell-1}]} \sim \mathcal{N}(\mathbf{x}_{i}^{[t_{\ell-1}]}, \sigma^2 \mi).
$
The parameters are estimated using Markov chain Monte Carlo methods. We also extend the initial model by associating $\{\mathbf{x}_{i}^{[t]}\}$ with outgoing edges and $\{\mathbf{y}_{i}^{[t]}\}$ with incoming edges, respectively. 
{\color{black}
To obtain clustering results, similar to FASE, we aggregate each vertex's latent positions over all time points and apply $k$-means to the aggregated features $\{[\hat{\mathbf{x}}_{i}^{[t_1]} \mid \cdots \mid \hat{\mathbf{x}}_{i}^{[t_m]}]\}_{i \in [n]}$ and $\{[\hat{\mathbf{y}}_{i}^{[t_1]} \mid \cdots \mid \hat{\mathbf{y}}_{i}^{[t_m]}]\}_{i \in [n]}$.
}
%$k$-means clustering is applied to $\{\hat{\mathbf{x}}_{i}^{[t]}\}_{i \in [n]}$ and $\{\hat{\mathbf{y}}_{i}^{[t]}\}_{i \in [n]}$ to obtain the clustering results.
\end{sloppypar}

\begin{figure}[htbp]
    \centering
    \includegraphics[width=0.76\textwidth]{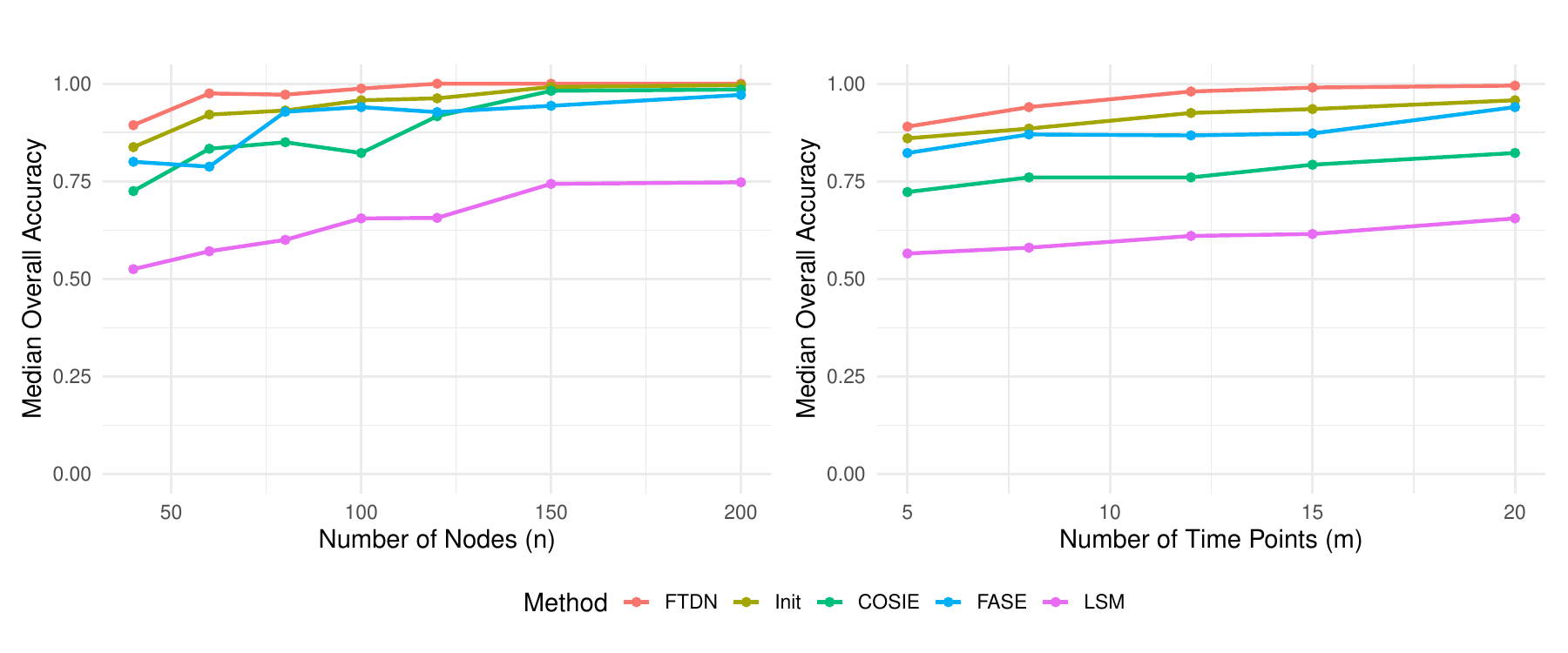}
    \caption{Empirical overall accuracy, measured as the average of the proportions of vertices correctly clustered for outgoing and incoming edges, using the FTDN and other methods under the following settings: (1) varying $n \in \{40, 60, 80, 100, 120, 150, 200\}$ while fixing $m = 20$, and (2) varying $m \in \{5, 8, 12, 15, 20\}$ while fixing $n = 100$. Additional details of the settings are provided in Section~\ref{sec:dynamic SBM}. The results are medians of $50$ independent Monte Carlo replicates.}
    \label{fig:simu_dynamic SBM}
\end{figure}

Figure~\ref{fig:simu_dynamic SBM} presents the results of varying the number of vertices $n$ and the number of time points $m$, demonstrating that accuracy improves as either $n$ or $m$ increases. 
Note that COSIE and the initialization of Algorithm~\ref{algorithm_rkhs} do not utilize temporal ordering or smoothness. 
LSM assumes an iterative dependence of parameters between consecutive time points, rather than leveraging temporal smoothness. 
FASE utilizes temporal smoothness, while FTDN further separates the time-invariant components. 
As illustrated in Figure~\ref{fig:simu_dynamic SBM}, FTDN outperforms other methods in accuracy.

\section{Real Data Experiments}\label{sec:real}

In this section, we use two real datasets, the Citi Bike trip data and the international food trade data, to demonstrate the performance of our model and algorithm on single dynamic networks and dynamic multilayer networks, respectively.

\subsection{Citi Bike trip dynamic network} \label{sec:citi}

Citi Bike is a public bike-sharing system serving New York City and its surrounding areas. The Citi Bike trip history data includes ride activities across the entire service area, providing details such as the starting and ending stations of trips and their timestamps. The data can be downloaded from \href{https://citibikenyc.com/system-data}{https://citibikenyc.com/system-data}.

\begin{figure}[htbp]
    \centering
    \includegraphics[width=0.35\textwidth]{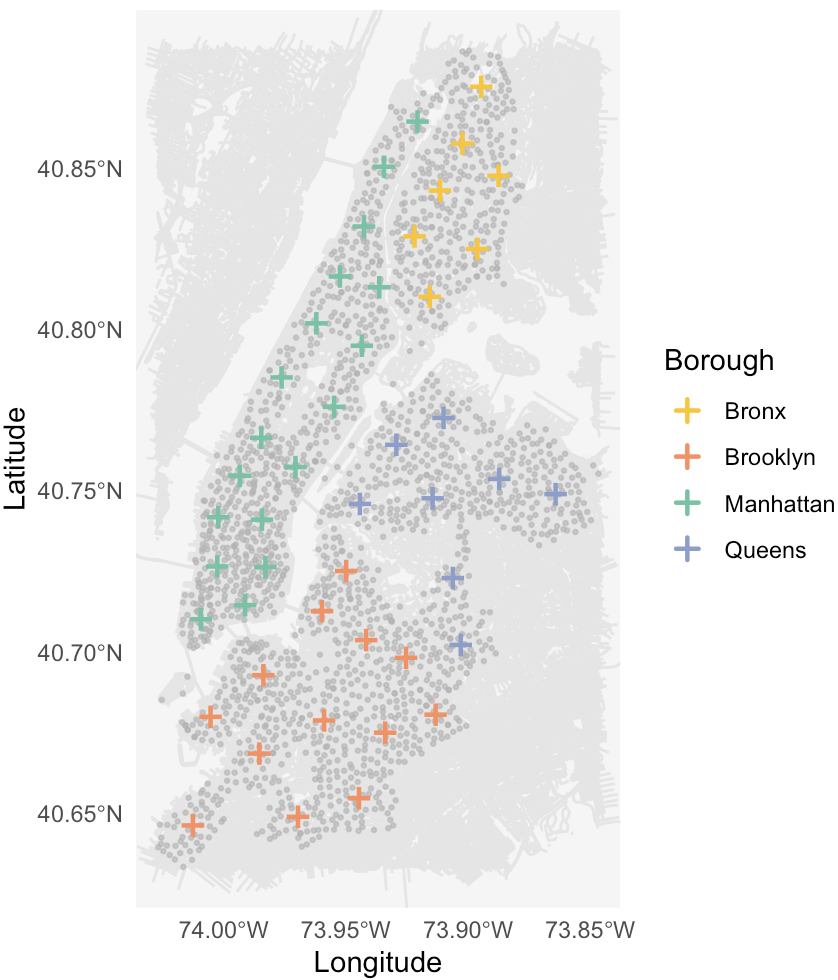}
    \caption{The $k$-means clustering results for the $2240$ Citi Bike stations in New York City, grouped into 46 clusters. The colored crosses represent the cluster centers, with the four boroughs, Bronx, Brooklyn, Manhattan, and Queens, represented by different colors. The smaller grey dots represent the original stations.}
    \label{fig:real_citi_borough}
\end{figure}

We analyze trip data from April 22, 2025, at 00:00 to April 25, 2025, at 00:00, covering three full days (April 22, 23, and 24), focusing on the $2240$ stations located in the four boroughs of New York City: Bronx, Brooklyn, Manhattan, and Queens. 
Due to the sparsity of the connections between the original $2240$ stations, we apply $k$-means clustering to group the stations into $46$ clusters, as shown in Figure~\ref{fig:real_citi_borough}, treating these clusters as the $n = 46$ vertices of the dynamic network. 
We divide the three-day period into two-hour intervals, resulting in $m = 36$ time points, with each interval serving as a time point. For the $n \times n$ adjacency matrices $\{\ma^{[t]}\}_{t=1}^{m}$, if a trip starts at a station in cluster $i$ and ends at a station in cluster $j$ within the $t$-th time interval, we set $\ma^{[t]}_{i,j} = 1$.

\begin{figure*}[htbp!]
    \centering
    \includegraphics[width=0.4\textwidth]{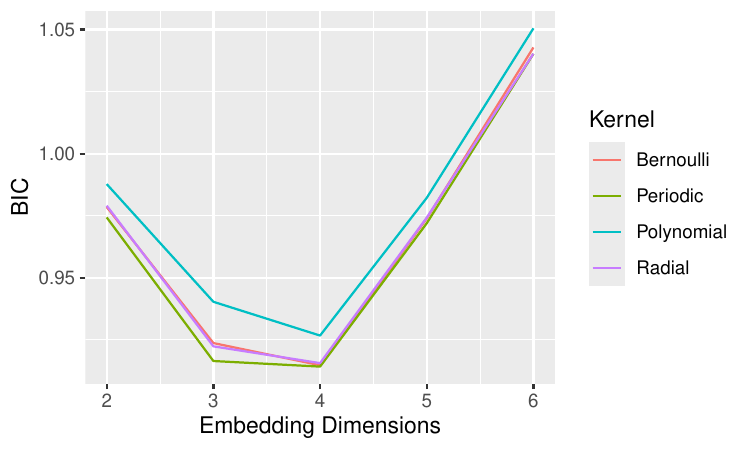}
    \includegraphics[width=0.4\textwidth]{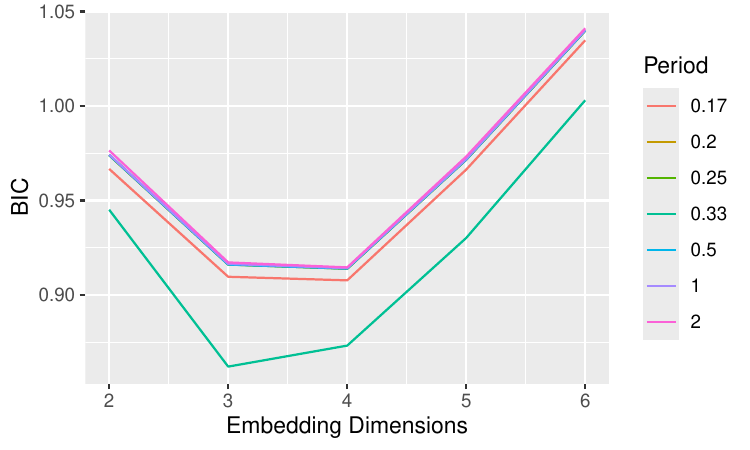}
    \caption{Kernel and kernel parameter selection by BIC. }\label{fig_bic}
\end{figure*}

%{\color{black} 
When applying Algorithm~\ref{algorithm_rkhs}, we first determine the kernel function by comparing the BIC among Bernoulli, polynomial, radial, and periodic kernels with some default kernel parameters across different embedding dimensions $d$ and then select the periodic kernel, which yields the smallest BIC at almost all $d$ (left panel in Figure \ref{fig_bic}). The choice of the kernel period is determined by comparing the BIC of different periods with different $d$ (right panel in Figure \ref{fig_bic}), and we ultimately select $d=3$ with a period of $1/3$, which aligns well with the nature of the data, a three-day trip dataset. 
%}

\begin{figure}[htbp!]
    \centering
    \includegraphics[width=0.75\textwidth]{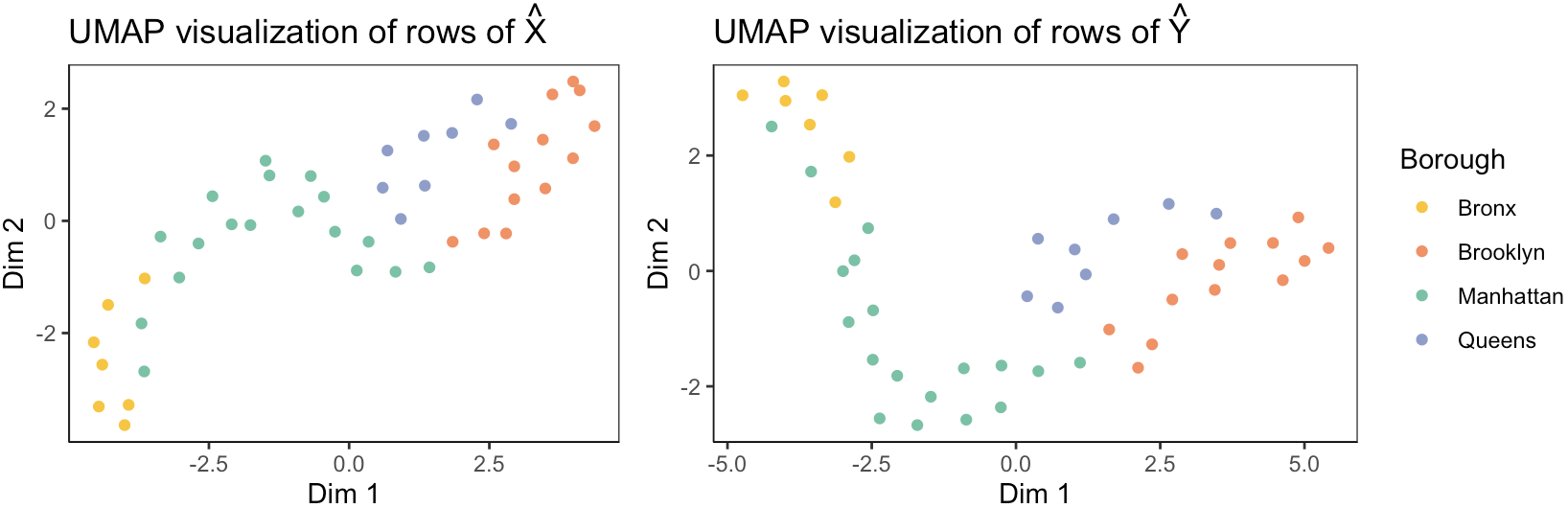}
    \caption{UMAP visualization of the rows of $\hat{\mx}$ and $\hat{\my}$, with colors indicating the boroughs of the vertices.}
    \label{fig:real_citi_XY}
\end{figure}

After obtaining $\hat{\mx}$ and $\hat{\my}$ using Algorithm~\ref{algorithm_rkhs}, we apply UMAP to visualize their rows, which correspond to the vertices, in a $2$-dimensional space, respectively, as shown in Figure~\ref{fig:real_citi_XY}. Figure~\ref{fig:real_citi_XY} illustrates that the vertices within each borough are close together, and the vertices from different boroughs are generally separated. This indicates that both the outgoing embeddings and incoming embeddings are strongly related to geographic locations.
{\color{black}We also provide PCA-based visualization in Appendix~\ref{sec:citi_add}, which exhibits similar patterns.}

\begin{figure}[htbp]
    \centering
    \includegraphics[width=0.45\textwidth]{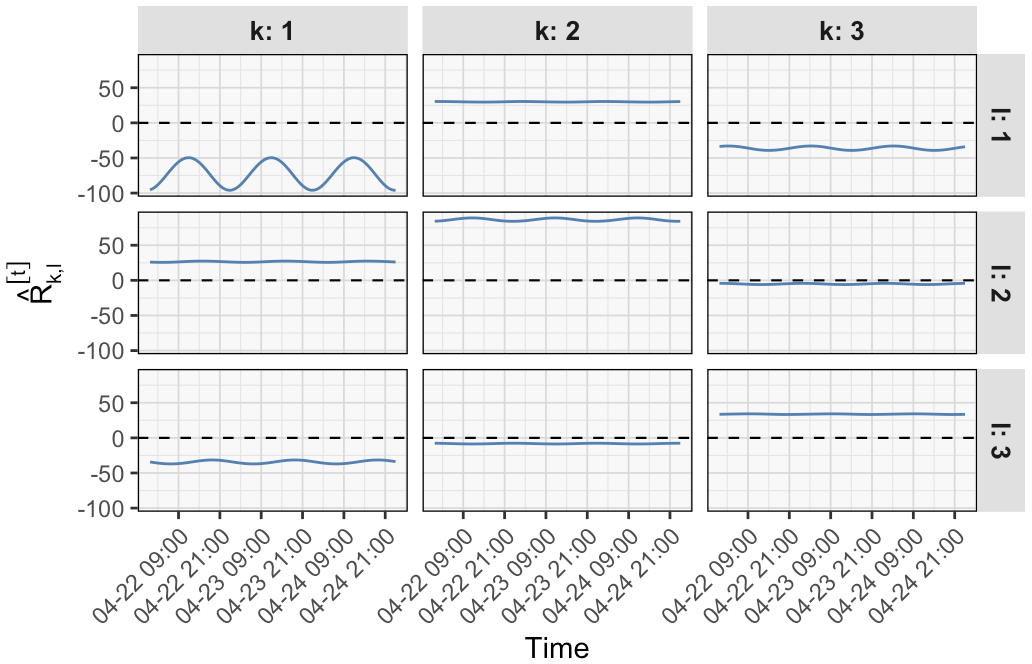}
    \hspace{0.01\textwidth}
    \includegraphics[width=0.5\textwidth]{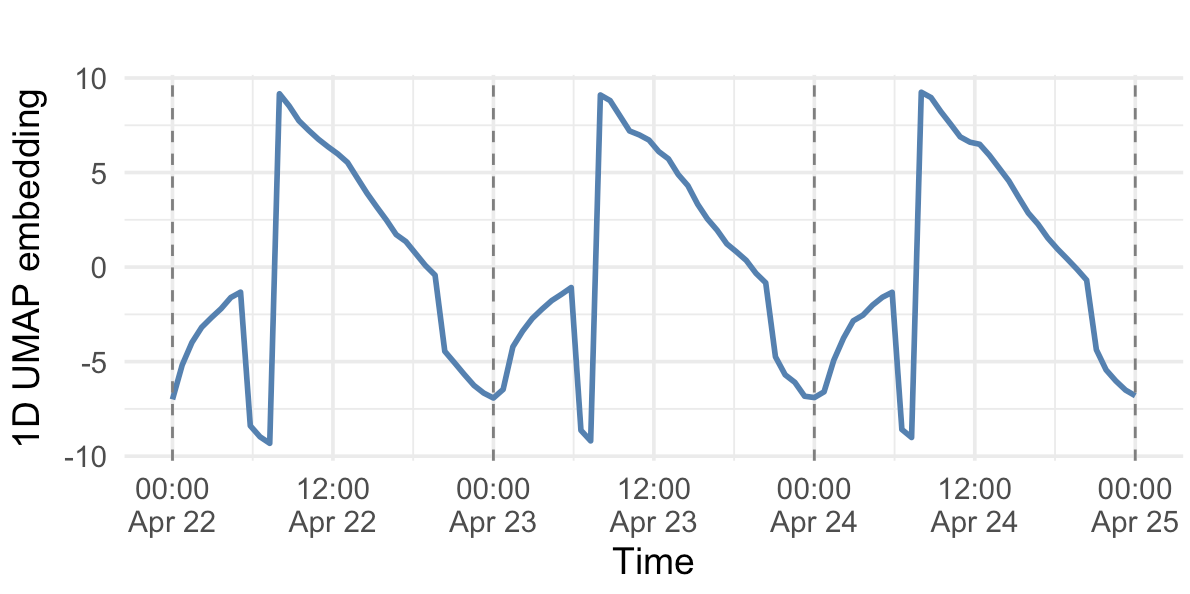}
    \caption{Temporal evolution of $\hat{\mr}^{[t]}$. The left panel shows the temporal trajectories of individual entries $(k,\ell)$ in $\hat{\mr}^{[t]}$. The right panel displays the $1$-dimensional UMAP embedding of $\hat{\mr}^{[t]}$ based on the Frobenius norm distances between time points to visualize the trajectory of the entire $\hat{\mr}^{[t]}$.}
    \label{fig:real_citi_R}
\end{figure}

From $\hat\mr^{[t]}$ obtained by Algorithm~\ref{algorithm_rkhs}, as mentioned in Section~\ref{sec:identifiability}, we obtain a $d^2$-dimensional trajectory representing the temporal evolution of the network's time-varying component. The left panel of Figure~\ref{fig:real_citi_R} shows the temporal trajectories of each entry in $\hat\mr^{[t]}$. Recall that the different entries of $\mr^{[t]}$ represent the interactions between the latent positions in different dimensions of $\mx$ and $\my$. The left panel of Figure~\ref{fig:real_citi_R} demonstrates that some dimensions exhibit interactions close to zero, while others are more pronounced, and some interactions show significant temporal changes, whereas others remain relatively stable. 
As described in Section~\ref{sec:identifiability}, we can further apply distance-based dimensionality reduction techniques using $\|\hat\mr^{[t]}-\hat\mr^{[t']}\|_F$ to visualize the network evolution. More specifically, we densely select $100$ time points to construct the distance matrix and then apply UMAP again %classical multidimensional scaling (CMDS)
 to reduce the dimensionality to one dimension, as shown in the right panel of Figure~\ref{fig:real_citi_R}. The visualization in the right panel of Figure~\ref{fig:real_citi_R} clearly reveals a three-day periodic pattern in the evolution of the trip network.

\subsection{Food trade dynamic multiple network}\label{sec:trade}

We analyze trade networks from 1993 to 2018 with layers representing different food and agricultural products. The data is provided by the Food and Agriculture Organization (FAO) of the United Nations and is available at \href{https://www.fao.org/faostat/en/#data/TM}{https://www.fao.org/faostat/en/\#data/TM}.

We construct a dynamic multilayer directed network, where the vertices represent trade entities (including countries and regions), each layer corresponds to a specific product, and the edges in each layer represent import-export trading relationships for that product. Each year is treated as a time point, resulting in $m=26$ time points. For the adjacency matrices $\{\ma^{[t]}_s\}$, an entry $\ma^{[t]}_{s,i,j} = 1$ is set if and only if entity $i$ exports product $s$ to entity $j$ in year $t$ with a trade value exceeding two hundred thousand USD.  
We focus on $K=10$ products as the layers:  
1) food preparations n.e.c.,  
2) crude organic material n.e.c.,  
3) wine,  
4) beer of barley (malted),  
5) undenatured ethyl alcohol (less than $80\%$ vol) and other spirits,  
6) other non-alcoholic caloric beverages,  
7) apples,  
8) maize (corn),  
9) wheat,  
10) potatoes.  
{\color{black}
From the FAO data, we selected these 10 products from those with the densest trading networks: two groups of similar products (beverages and crops) and two contrasting products (food preparations and organic material). This selection allows us to examine both the model's ability to cluster similar products and to separate distinct ones.
}
We finally extract the intersection of the largest connected components from the adjacency matrices of these products across all years, resulting in a dynamic multilayer network with $n=62$ entities as vertices.
Following a parameter selection process similar to Section~\ref{sec:citi}, we select the embedding dimension $d=3$ using the periodic kernel with period 1{\color{black}; details on kernel and dimension selection are provided in Appendix~\ref{sec:trade_add}.}

\begin{figure}[htbp]
    \centering
    \includegraphics[width=0.91\textwidth]{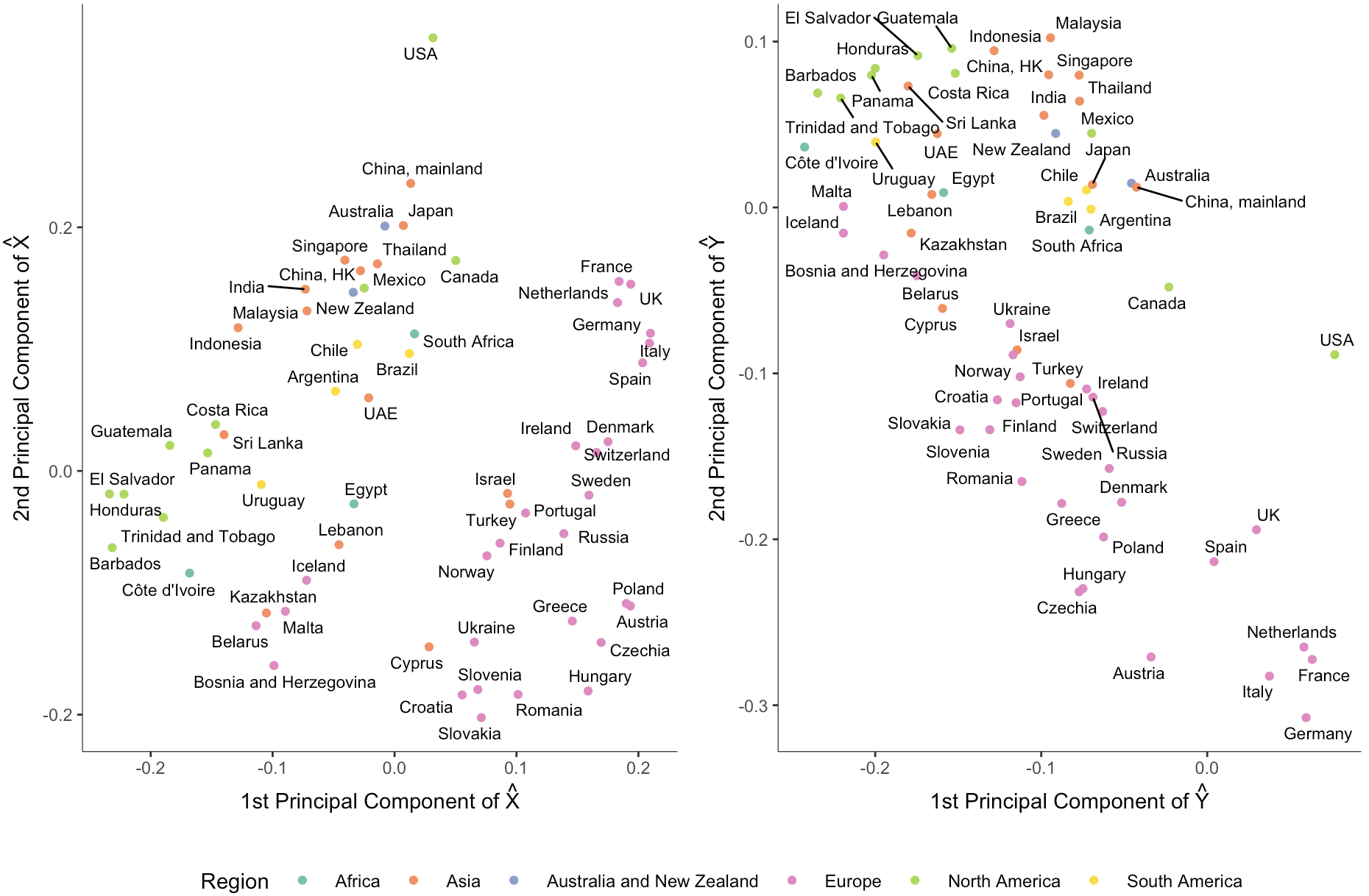}
    \caption{PCA visualization of the embeddings of trade entities from $\hat{\mx}$ and $\hat{\my}$ using the first two principal components. Each point represents a trade entity, with colors indicating its region.}
    \label{fig:real_trade_XY}
\end{figure}

After obtaining $\hat\mx$ and $\hat\my$, we visualize the entity export and import embeddings using the first two principal components in Figure~\ref{fig:real_trade_XY}. Figure~\ref{fig:real_trade_XY} shows that the embeddings of entities within the same region are generally close, reflecting the correlation between geographic proximity and trade patterns.
{\color{black}
We also provide UMAP visualization in Appendix~\ref{sec:trade_add}, which shows similar patterns.
}

\begin{figure}[htbp]
    \centering
    \includegraphics[width=0.32\textwidth]{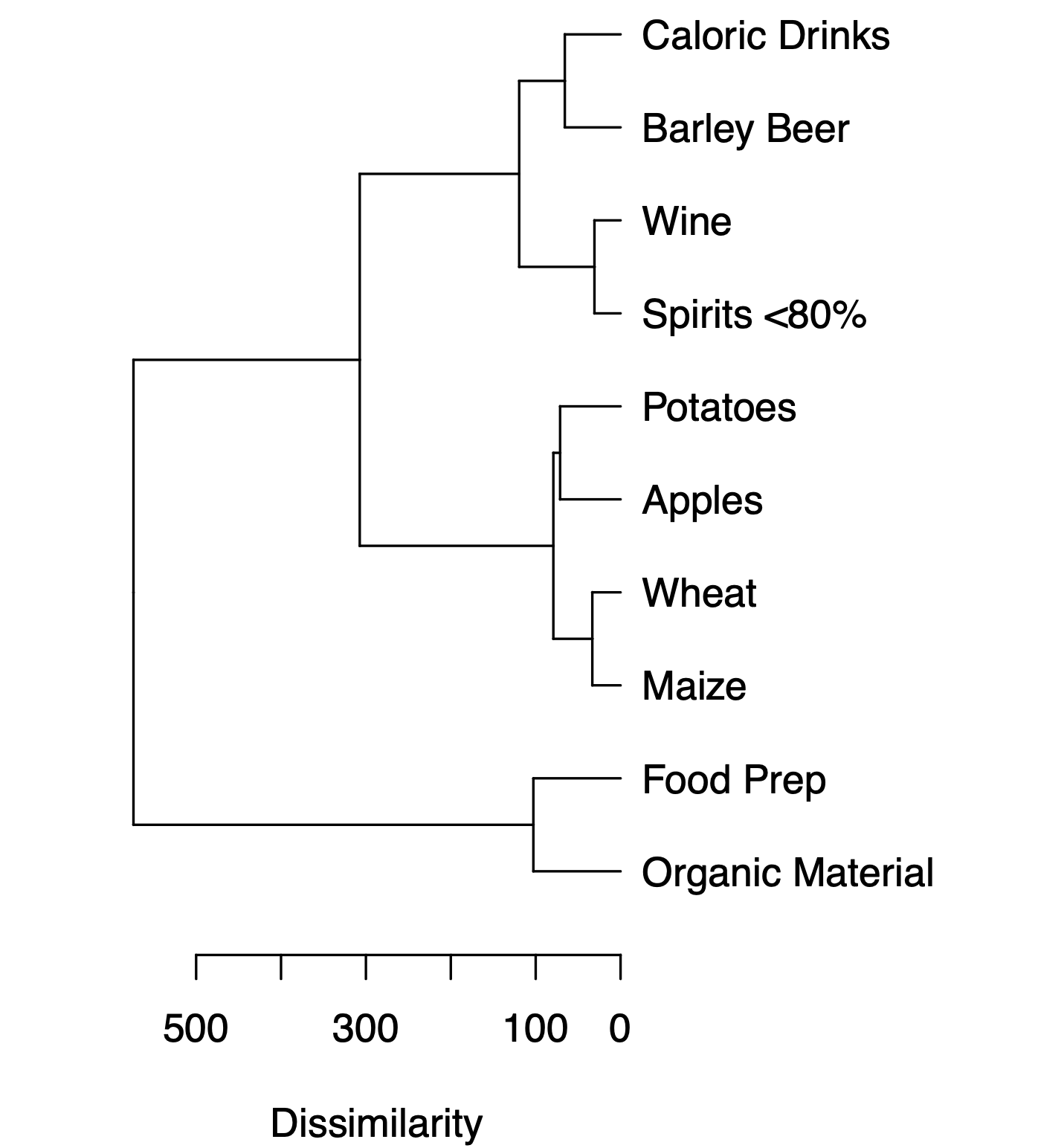}
    \includegraphics[width=0.67\textwidth]{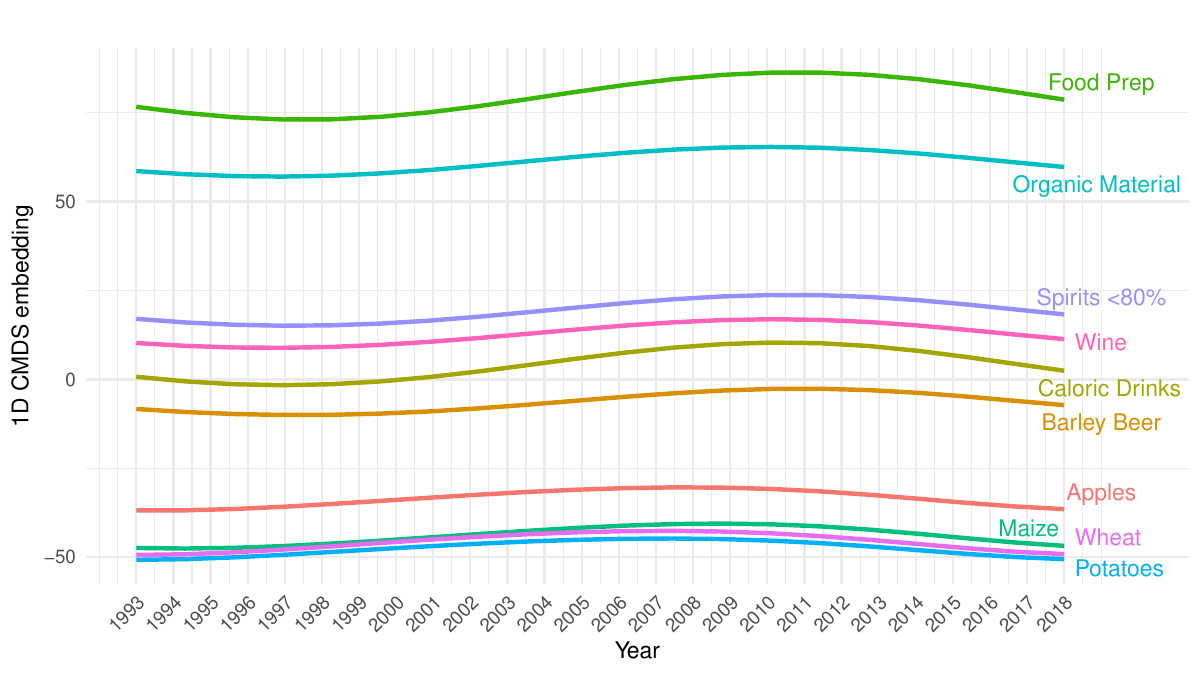}
    \caption{Results about temporal evolution of $\hat{\mr}^{[t]}_s$.
    Left panel shows hierarchical clustering of network evolution trajectories for different products, based on the trajectory-level Euclidean distances. 
     The left panel shows hierarchical clustering of layer trajectories. The right panel displays the 1-dimensional CMDS embedding of $\hat{\mr}^{[t]}_s$ based on the Frobenius norm distances between time points and/or layers to visualize the layer trajectories. %{\color{black} (use $\theta$ to do the hierarchical clustering instead of $\hat\mr$?)}
     }
    \label{fig:real_trade_R}
\end{figure}

The obtained $\hat\mr^{[t]}_s$ provides the trajectories of network evolution for all layers (products) in the same $d^2$-dimensional space. We evaluate the difference between these trajectories and analyze the similarity or dissimilarity of network evolution across different products, as mentioned in Section~\ref{sec:identifiability}. Specifically, we compute the trajectory-level Euclidean distances between the $\hat\mr^{[t]}_s$ trajectories for different product $s$ and perform hierarchical cluster analysis. The clustering results are presented in the left panel of Figure~\ref{fig:real_trade_R}, where four beverage products form one cluster, four crop products form another cluster, and food preparations and organic material cluster separately.
We also visualize the trajectories together by applying classical multidimensional scaling (CMDS) based on the distance $\|\hat\mr^{[t]}_s-\hat\mr^{[t']}_{s'}\|_F$ between different time points and/or network layers. The visualization is shown in the right panel of Figure~\ref{fig:real_trade_R}, where the trajectories exhibit a structure similar to the hierarchical clustering results.

\section{Extensions}

\label{sec:extension}

In this section, we introduce several extensions to our model and algorithm. In Section~\ref{sec:missing}, we explain how the algorithm can be adapted to handle missing data and even unaligned observations. In Section~\ref{sec:model with layer interaction}, we generalize MFTDN to account for more complex inter-layer dependencies, where the generalized model can be regarded as a generalized functional tensor Tucker decomposition. And in Section~\ref{sec:gtucker}, we demonstrate how the RKHS estimation proposed in Section~\ref{sec:estimation} can be extended to a broader setting of generalized functional tensor Tucker decomposition. Finally, in Section~\ref{sec:model with smooth s}, we discuss the extended MFTDN model that can incorporate smoothness in both temporal and layer dimensions, along with its corresponding estimation method.

\subsection{Missing data and unaligned observations}\label{sec:missing}

{\color{black}
%The estimation approach described in Section~\ref{sec:estimation} can be adapted to handle scenarios with missing data. 
%Specifically, if some edge connections are missing for certain observed time points or layers, we can adjust Eq.~\eqref{} in our algorithm by ... to incorporate these varying observation sets across different $(s, t)$.
}

%{\color{black}
The Algorithm \ref{algorithm_rkhs} described in Section~\ref{sec:estimation} can also be further extended to handle even unaligned data. 
Since the temporal dimension is assumed to be continuous, it is possible to observe partial edge connections for partial layers at each arbitrary time point.
Solving the functional tensor problem for such unaligned data is feasible with appropriate modifications, as demonstrated in \cite{tang2024tensor, larsen2024tensor}. Specifically, note that the loss function defined in \eqref{eq_MLE} is a summation of the pairwise loss across all observations. Hence, in the case of missing or aligned data, we can rewrite it as 
\beas\label{eq_MLE_unaligned}
     \argmin_{\mx, \my, \mr} \sum_{i,j\in[n]; s \in [K]} \sum_{t \in \mathcal{T}_{i,j,s}} f\(\ma^{[t]}_{s,i,j}, \sum_{k \in [d]} \sum_{\ell \in [d]}(\mathbf{x}_i)_k \, \mr^{[t]}_{s, k,\ell} \, (\mathbf{y}_j)_\ell\). 
\eeas
Here, $\mathcal{T}_{i,j,s}$ is the set of time points when the edge $\ma_{s,i,j}$ is observed. Proposition \ref{proposition_representer_thm} still holds with this modification, and hence we can discretize the functionals in $\mr$ and solve it by the gradient method. 
%}

In cases of missing data, where for each $(s, t)$ we observe an adjacency matrix $\ma^{[t]}_s$ that may contain missing entries, we can still achieve an effective initialization. If the missing entries are approximately uniformly distributed and relatively sparse, we can simply fill the missing entries with $\frac{1}{2}$ and follow the procedures described in Section~\ref{sec:initial} to obtain $\hat{\mx}^{[t]}_s$ and $\hat{\my}^{[t]}_s$ in Step~1. For more complex missing patterns, such as block-wise missing adjacency matrices, methods proposed in \cite{zheng2024chain, bishop2014deterministic} can be adapted to estimate $\hat{\mx}^{[t]}_s$ and $\hat{\my}^{[t]}_s$. 
However, if the observations are completely unaligned, it is no longer feasible to estimate the initial $\hat{\mx}$ and $\hat{\my}$ using the spectral method. In this case, a random initialization might be considered. 

\subsection{Generalized functional tensor Tucker decomposition and RKHS estimation}\label{sec:gtucker}

The RKHS framework for solving MLE in Section~\ref{sec:estimation} can be generalized to a broader setting of tensor decomposition. 
Assume that for some functional tensor $\ma^{[t]} \in \RR^{p_1 \times \cdots \times p_D \times [0,1]}$, we observe its values $\{\ma_{i_1 , \ldots , i_D}^{[t]}\}_{i_k \in [p_k], k \in [D], t\in \mathcal{T}}$. 
We are interested in finding a low Tucker rank functional tensor 
\bea\label{eq_functional_tucker_model}
\mpp^{[t]} = \mr^{[t]} \times_{k \in [D]} \mx_k\in \RR^{p_1 \times \cdots \times p_D \times [0,1]}
\eea
where $\mr^{[t]} \in \RR^{r_1 \times \cdots \times r_D \times [0,1]}$ is a functional core tensor, and $\mx_k \in \mathcal{O}_{p_k \times r_k}$ are orthonormal factor matrices,
to minimize the empirical loss
\bea\label{eq_general_loss_1}
     \argmin_{\substack{\mr \in \RR^{r_1 \times \cdots \times r_D \times [0,1]} \\ \mx_k \in \mathcal{O}_{p_k \times r_k}}} \sum_{i_k \in [p_k], k \in [D], t\in \mathcal{T}} f\(\ma_{i_1 , \ldots , i_D}^{[t]}, (\mr^{[t]} \times_{k \in [D]} \mx_k)_{i_1 , \ldots , i_D} \)
\eea
for some pairwise loss function $f$. To apply the RKHS framework, we assume $\mr_{i_1 , \ldots , i_D}$ is in some RKHS $\mathcal{H}$ with kernel $\mathbb{K}(\cdot, \cdot)$. With some smoothness and overfitting constraints, we rewrite \eqref{eq_general_loss_1} as 
\bea\label{eq_general_loss_2}
     &&\argmin_{{\substack{\mx_k \in \mathbb{R}^{p_k \times r_k}, \mr_{i_1 , \ldots , i_D} \in \mathcal{H} \\ \sum_{j_k \in [r_k]; k \in [D]} \(\|\mr_{j_1 , \ldots , j_D}\|_{\mathcal{H}} \|(\mx_1)_{:j_k}\| \ldots \|(\mx_D)_{:j_k}\|\)\leq C}}} \notag\\
     && 
     \sum_{i_k \in [p_k], k \in [D], t\in \mathcal{T}} f\(\ma_{i_1 , \ldots , i_D}^{[t]}, (\mr^{[t]} \times_{k \in [D]} \mx_k)_{i_1 , \ldots , i_D} \). 
\eea
A generalization of Proposition \ref{proposition_representer_thm} still applies if $f(x,y)$ is convex and Gateaux-differentiable, i.e., \eqref{eq_general_loss_2} admits a solution where $\hat \mr_{i_1 , \ldots , i_D}$ can be represented as 
\[
    \hat \mr_{i_1 , \ldots , i_D}^{[t]} = \sum_{h \in \mathcal{T}} \theta_{i_1 , \ldots , i_D, h} \mathbb{K}(t, h).
\]
Hence, we can solve \eqref{eq_general_loss_2} by evaluating
\bea\label{eq_general_loss_3}
     &&\argmin_{{\substack{\mx_k \in \mathbb{R}^{p_k \times r_k}, \theta \in \RR^{r_1\ldots r_D} \\ \sum_{j_k \in [r_k]; k \in [D]} \(\|\mr_{j_1 , \ldots , j_D}\|_{\mathcal{H}} \prod_{h \in [D]} \|(\mx_h)_{:j_k}\| \)\leq C}}} \notag\\
     &&\sum_{i_k \in [p_k], k \in [D], t\in \mathcal{T}} f\(\ma_{i_1 , \ldots , i_D}^{[t]}, \sum_{j_k \in [r_k]; k \in [D]} (\mx_k)_{i_1, j_1} \ldots (\mx_k)_{i_D, j_D} \sum_{h \in \mathcal{T}} \theta_{j_1 , \ldots , j_D, h} \mathbb{K}(t, h)\). 
\eea
Now, \eqref{eq_general_loss_3} is a finite-dimensional optimization problem and hence can be solved by an algorithm similar to Algorithm \ref{algorithm_rkhs}.   

Similar to the settings in \cite{tang2024tensor, larsen2024tensor}, the set of observation times $\mathcal{T}$ is allowed to vary with the indices $i_1, \ldots, i_D$, meaning we observe $\{\ma_{i_1 , \ldots , i_D}^{[t]}\}_{i_k \in [p_k], k \in [D], t\in \mathcal{T}_{i_1, \ldots, i_D}}$. This flexibility allows our framework to accommodate unaligned or irregularly sampled functional tensor observations. Consequently, the model described in Section~\ref{sec:model} naturally handles cases where there are missing edges at some observation times, or even different edges are observed at different time points. 

The decomposition in \eqref{eq_functional_tucker_model} was also studied in the recent work of \cite{hanTensorFactorModel2024} and references therein, where it is interpreted as a tensor time series model and addressed via iterative orthogonal projections. In contrast, our proposed approach is more general, as it accommodates a broader class of loss functions and distributions of noises, beyond the $\ell_2$ loss and Gaussian noise considered in \cite{hanTensorFactorModel2024}.

\subsection{Generalized functional tensor dynamic multilayer network model with layer interaction}\label{sec:model with layer interaction}

In the MFTDN model, we assume that all network layers share common invariant subspaces. To further capture more flexible and structured inter-layer relationships, we extend MFTDN to the following generalized functional tensor multilayer network model. Specifically, we model the edge probability functional tensor $\tilde\mpp \in \mathbb{R}^{K \times n \times n \times [0,1]}$ as  
$$
\tilde \mpp^{[t]} = \mr^{[t]} \times_2 \mx \times_3 \my \times_1 \mz,
$$
where  
$\mx, \my \in \mathbb{R}^{n \times d}$ are still the node-specific outgoing and incoming embeddings,  
$\mr^{[t]} \in \mathbb{R}^{M \times d \times d \times [0,1]}$ is a time-varying core tensor representing $M$ latent interaction modes, assumed to vary smoothly over time $t$,  
and $\mz \in \mathbb{R}^{K \times M}$ encodes how each of the $K$ layers expresses the $M$ latent interaction components. {\red This formulation fits into the generalized functional tensor Tucker decomposition in \eqref{eq_functional_tucker_model}. }

This formulation generalizes MFTDN by introducing additional flexibility, where setting $M = K$ and $\mz = \mathbf{I}_K$ recovers the structure in MFTDN. 
This added flexibility enables the generalized model to capture more complex inter-layer dependencies within the multilayer network. 
Specifically, $\mr^{[t]}$ can be interpreted as $M$ trajectories in the $d^2$-dimensional space. By applying $\mz$ to the first mode of $\mr^{[t]}$, these $M$ trajectories are linearly combined to generate the $K$ interaction trajectories associated with the $K$ network layers. Hence, the $K$ rows of $\mz$ reflect the structural relationships among the $K$ network layers.

Nevertheless, the MFTDN model, with its clearer structure, remains more tractable for some inference tasks, as discussed in Section~\ref{sec:identifiability}. For example, MFTDN enables a more straightforward and interpretable visualization of the evolution patterns across different network layers.

This generalized functional tensor model can be estimated using the algorithm developed for generalized Tucker decomposition; see Section~\ref{sec:gtucker} for details.

%{\color{black}(how to choose d and M?
%run trading network data?)}

\subsection{Extended MFTDN with smoothness in both temporal and layer dimensions}\label{sec:model with smooth s}

In the current MFTDN framework, we assume that the networks $\{\ma^{[t]}_s\}$ are temporally ordered along the time dimension $t$, while no specific order is defined between layers. 
In certain scenarios, the layers may also follow a well-defined order. For example, in brain dynamic multilayer networks where the layers correspond to different frequency bands such as the Delta layer ($0.5-4 \, \mathrm{Hz}$), Theta layer ($4-8 \, \mathrm{Hz}$), Alpha layer ($8-13 \, \mathrm{Hz}$), Beta layer ($13-30 \, \mathrm{Hz}$), low Gamma layer ($30-50 \, \mathrm{Hz}$), and high Gamma layer ($50-100 \, \mathrm{Hz}$) \citep{de2017multilayer,buldu2018frequency}. In this case, the layers exhibit a natural order, and smoothness is expected across layers that correspond to similar frequency bands.
Thus, for $\mr^{[t]}_s$, in addition to smoothness along the $t$-dimension, we also assume smoothness along the $s$-dimension. 

%{\color{black}
The extended MFTDN model, which incorporates the assumption that $\mr^{[t]}_s$ is smooth in both the time dimension $t$ and the layer dimension $s$, can still be estimated using the RKHS-based method. Now we can rewrite $\mr^{[t]}_s$ as $\mr^{[t, s]}$. The Proposition \ref{proposition_representer_thm} still holds, so that the optimizer $\mr^{[t, s]}$ for fixed $\mx$ and $\my$ can be written as 
\beas
        \hat \mr^{[t, s]}_{k,\ell} = \sum_{q \in [K], h \in \mathcal{T}} \theta_{k,\ell, q, h} \mathbb{K}((t, s), (h, q)). 
\eeas

%Q: In modeling the two-dimensional function $\mr^{[s,t]}$, is it necessary to use a separable kernel for smoothing? Or can one flexibly choose between separable and non-separable kernels, with the former offering computational advantages?

%Kernel separability here does not matter. I dont think we need to talk about computational advantage because we did not mention the computational cost in this work. The answer is yes, but not significant. With separability, the number of estimated parameters does not change, but at very least the storing size of the kernel matrix reduces from $|\mathcal{S}|^2 |\mathcal{T}|^2$ to $2(|\mathcal{S}|^2 + |\mathcal{T}|^2)$, which will also reduce time cost when it is used in calculation. 
%}

%\subsection{Generalized eigenscaling Model and functional tensor CP decomposition}

\section{Discussion}\label{sec:discussion}

The method proposed in this paper can be extended to broader networks that exhibit both smooth evolution and abrupt changes. The MFTDN and the corresponding algorithm are based on the assumption that networks evolve smoothly over time. This assumption covers many scenarios where networks change gradually, and leveraging this smoothness enables more accurate estimation. For cases where networks do not remain smooth across all time points, i.e., when there exist change points at certain time points, we can first apply change point detection algorithms \citep{wang2021optimal, ranshous2015anomaly} to identify these discontinuities, and then assume continuity within each detected segment to adapt the method.

The MFTDN model can also be utilized for offline learning, expanding its range of applications. For example, if we have used a large dataset of brain dynamic networks to obtain $\hat{\mx}$, $\hat{\my}$, and to learn the $\{\hat{\mr}^{[t]}\}$ associated with specific diseases, we can efficiently classify new brain dynamic networks. Specifically, given a new brain dynamic network, based on the pre-trained $\hat{\mx}$ and $\hat{\my}$, we can solve the convex optimization problem in Step~2 of Section~\ref{sec:initial} to compute the corresponding $\hat{\mr}^{\text{new}[t]}$, and then compare $\hat{\mr}^{\text{new}[t]}$ with the existing set $\{\hat{\mr}^{[t]}\}$ associated with known diseases. This approach eliminates the need for repetitive training on large datasets and safeguards the privacy of the training data.

Finally, the MFTDN model is highly flexible and can be extended to scenarios involving arbitrary dimensions. For instance, as discussed in Section~\ref{sec:model with smooth s}, consider brain dynamic multilayer networks where the layers correspond to different frequency bands, where $\mr$ has two dimensions, time and frequency bands. Furthermore, if we consider such brain dynamic multilayer networks for multiple subjects, $\mr$ can include a third dimension corresponding to subjects. In this example, $\mr$ has two smooth dimensions (time and frequency bands) and one unordered dimension (subjects). The MFTDN model can be easily adapted to specific applications, depending on the needs of the problem at hand.

\newpage

\bibliographystyle{unsrtnat}
\bibliography{ref_full}

\newpage

\begin{appendices}

% \begin{center}%
%     {\LARGE Supplementary Material for ``A Functional Tensor Model for Dynamic Multilayer Networks with Common Invariant Subspaces and the RKHS Estimation"\par}%
%   \end{center}

\counterwithin{figure}{section}

\counterwithin{propositi}{section}

% \begin{center}%
%     {\LARGE Supplementary Material for ``A Functional Tensor Model for Dynamic Multilayer Networks with Common Invariant Subspaces and the RKHS Estimation"\par}%
%   \end{center}

{\color{black}
\section{Additional Discussion}

\subsection{Additional discussion for FTDN}\label{app:add discussion for FTDN}

As mentioned in Section~\ref{sec:single}, beyond the standard dynamic SBM with fixed community assignments, the FTDN model can capture smooth community evolution, including transitions, splits, and mergers, when such patterns can be represented by a generalized dynamic SBM where $\mathbf{B}^{[t]}$ is allowed to be rank-deficient at some time points. As illustrated in the transition example described in Section~\ref{sec:single}, $\mathbf{B}^{[t]}$ has rank 2 at both the initial and final time points (when the transitioning community behaves identically to the source and target communities, respectively), while achieving full rank 3 during the intermediate transition period. 
{\color{black}This flexibility makes FTDN particularly valuable in scenarios involving a small number of merging, splitting, or transition events where many nodes change their membership simultaneously.}

The induced rank $d$ of the FTDN representation (i.e., the dimension of $\mathbf{R}^{[t]}$, or equivalently, the dimension of $\mathbf{B}^{[t]}$ in the corresponding generalized dynamic SBM) equals the size of the finest partition of node behaviors across time. For the transition example in Section~\ref{sec:single}, where we have two communities and some nodes move from one community to another, the induced rank is $d = 3$. Other examples: when a community splits into two communities, the induced rank can be $d = 2$; when three communities merge into one, the induced rank can be $d = 3$.

\subsection{Additional discussion for MFTDN}\label{app:add discussion for MFTDN}

As mentioned in Section~\ref{sec:multilayer}, the MFTDN model can accommodate dynamic multilayer SBMs even when community assignments differ across layers, and corresponds to a generalized dynamic multilayer SBM where layer- and time-specific block probability matrices $\mathbf{B}^{[t]}_s$ can be rank-deficient. 
For example, consider 2 layers where Layer 1 has 3 communities and Layer 2 has 3 communities formed by: community 1 as a merger of Layer 1's communities 1 and 2, and communities 2 and 3 as splits of Layer 1's community 3. Thus the finest partition of vertices yields 4 distinct communities, where communities 3 and 4 behave identically in Layer 1 and communities 1 and 2 behave identically in Layer 2. This gives an MFTDN induced rank of 4.

Notably, the non-trivial improvement from jointly modeling different layers compared to modeling each layer separately as a single dynamic SBM relies on the existence of common structure across layers. Such common structure is present, for example, in the example above, where although the community assignments differ between the two layers, the community structure still exhibits shared patterns through splitting and merging. If layers are completely independent with no structural relationships, then jointly modeling them by simply increasing dimensionality of the model would not yield non-trivial improvement over separate models.

}

{\red
\subsection{Convergence of Algorithm \ref{algorithm_rkhs}} \label{sec:convergence}

A rigorous projected-gradient guarantee for Algorithm \ref{algorithm_rkhs} can be obtained if the normalization step is replaced by the exact Euclidean projection onto a convex constraint set. 
This can be done by normalizing $\mx, \my$ and $\theta$, respectively. 
The normalization step in Algorithm~\ref{algorithm_rkhs} should be viewed as a practical stabilization heuristic. 

\begin{propositi}\label{proposition_convergence}
Let $z=(\mx,\my,\theta)$ denote the finite-dimensional parameter vector in
\eqref{eq_optimization_problem_2}, and let $\mathcal C$ be a nonempty, closed, convex feasible set.
Assume that $F(z)$ is continuously differentiable, bounded below on $\mathcal C$,
and has $L$-Lipschitz gradient on a neighborhood of $\mathcal C$.
Consider the projected gradient descent iteration
\[
z^{r+1}=\Pi_{\mathcal C}\bigl(z^r-\alpha \nabla F(z^r)\bigr),
\qquad 0<\alpha<1/L.
\]
Then
\[
F(z^{r+1})
\le
F(z^r)
-
\left(\frac{1}{2\alpha}-\frac{L}{2}\right)\|z^{r+1}-z^r\|^2.
\]
Consequently, $\{F(z^r)\}$ is nonincreasing and convergent, and
$\|z^{r+1}-z^r\|\to 0$.
Moreover, every accumulation point of $\{z^r\}$ is a first-order stationary point
of the constrained problem
\[
\min_{z\in\mathcal C} F(z).
\]
\end{propositi}
\begin{proof}
Let $\Delta^r := z^{r+1}-z^r$. Since $z^{r+1}$ is the Euclidean projection of
$z^r-\alpha \nabla F(z^r)$ onto $\mathcal C$, the projection optimality condition gives
\[
\left\langle
z^{r+1}-\bigl(z^r-\alpha \nabla F(z^r)\bigr),\, z-z^{r+1}
\right\rangle \ge 0,
\qquad \forall z\in\mathcal C.
\]
Taking $z=z^r\in\mathcal C$ yields
\[
\langle \nabla F(z^r), \Delta^r\rangle
\le
-\frac{1}{\alpha}\|\Delta^r\|^2.
\]
Since $\nabla F$ is $L$-Lipschitz, the descent lemma gives
\[
F(z^{r+1})
\le
F(z^r) + \langle \nabla F(z^r), \Delta^r\rangle
+\frac{L}{2}\|\Delta^r\|^2.
\]
Combining the last two displays,
\[
F(z^{r+1})
\le
F(z^r)-\left(\frac{1}{\alpha}-\frac{L}{2}\right)\|\Delta^r\|^2.
\]
Because $\alpha<1/L$, we have $\frac{1}{\alpha}-\frac{L}{2}\ge \frac{1}{2\alpha}$, and
hence
\[
F(z^{r+1})
\le
F(z^r)-\frac{1}{2\alpha}\|\Delta^r\|^2.
\]
Therefore $\{F(z^r)\}$ is nonincreasing. Since $F$ is bounded below on $\mathcal C$,
the sequence $\{F(z^r)\}$ converges, and summing the above inequality over $r$ gives
\[
\sum_{r=0}^\infty \|z^{r+1}-z^r\|^2 < \infty,
\]
so $\|z^{r+1}-z^r\|\to 0$.

Now let $z^{r_j}\to z^\ast$ be an accumulation point. Since $\Delta^{r_j}\to 0$, we also
have $z^{r_j+1}\to z^\ast$. The projection optimality condition can be rewritten as
\[
\left\langle
\nabla F(z^r)+\frac{1}{\alpha}(z^{r+1}-z^r),\, z-z^{r+1}
\right\rangle \ge 0,
\qquad \forall z\in\mathcal C.
\]
Passing to the limit along $r=r_j$ and using the continuity of $\nabla F$ gives
\[
\langle \nabla F(z^\ast), z-z^\ast\rangle \ge 0,
\qquad \forall z\in\mathcal C.
\]
This is exactly the first-order stationarity condition for the constrained problem.
\end{proof}

\begin{propositi}\label{proposition_convergence_2}
Assume that $C=\infty$, and let $z=(\mx,\my,\theta)$.
Suppose $F(z)$ is continuously differentiable, bounded below, and $L$-smooth
on a set containing all iterates generated by Algorithm~\ref{algorithm_rkhs}.
If $0<\alpha<1/L$, then
\[
F(z^{r+1})
\le
F(z^r)
-
\left(\frac{1}{\alpha}-\frac{L}{2}\right)\|z^{r+1}-z^r\|^2.
\]
Hence $\{F(z^r)\}$ is nonincreasing and convergent, and every accumulation point
of $\{z^r\}$ is a stationary point of $F$.
\end{propositi}

\begin{proof}
Let $\Delta^r:=z^{r+1}-z^r=-\alpha \nabla F(z^r)$. By the descent lemma,
\[
F(z^{r+1})
\le
F(z^r)+\langle \nabla F(z^r), \Delta^r\rangle
+\frac{L}{2}\|\Delta^r\|^2.
\]
Using $\Delta^r=-\alpha \nabla F(z^r)$,
\[
\langle \nabla F(z^r), \Delta^r\rangle
=
-\alpha \|\nabla F(z^r)\|^2
=
-\frac{1}{\alpha}\|\Delta^r\|^2.
\]
Hence
\[
F(z^{r+1})
\le
F(z^r)-\left(\frac{1}{\alpha}-\frac{L}{2}\right)\|\Delta^r\|^2.
\]
Since $\alpha<1/L$, we have $\frac{1}{\alpha}-\frac{L}{2}\ge \frac{1}{2\alpha}$, and therefore
\[
F(z^{r+1})
\le
F(z^r)-\frac{1}{2\alpha}\|z^{r+1}-z^r\|^2.
\]
Thus $\{F(z^r)\}$ is nonincreasing. Because $F$ is bounded below, the sequence
$\{F(z^r)\}$ converges, and summing the inequality over $r$ yields
\[
\sum_{r=0}^\infty \|z^{r+1}-z^r\|^2 < \infty,
\]
so $\|z^{r+1}-z^r\|\to 0$. Since
\[
\nabla F(z^r) = -\frac{1}{\alpha}(z^{r+1}-z^r),
\]
we conclude that $\|\nabla F(z^r)\|\to 0$.

Finally, let $z^{r_j}\to z^\ast$ be any accumulation point. By continuity of
$\nabla F$,
\[
\nabla F(z^\ast)=\lim_{j\to\infty}\nabla F(z^{r_j})=0.
\]
Therefore $z^\ast$ is a stationary point of $F$.
\end{proof}

}

{\color{black}
\subsection{Additional discussion on asymptotic results}\label{sec:asym}

The primary challenge in establishing asymptotic theoretical results for our estimators of the MFTDN model arises from the involvement of (functional) tensor decomposition, which remains largely unresolved in the literature. In fact, even in existing work on (functional) tensor decomposition, rigorous and complete theoretical results are rarely provided (e.g., \cite{kolda2009tensor, tang2024tensor}). 

Despite the fact that these challenges from functional tensor decomposition are difficult to overcome, we can still provide some insights into the asymptotic behavior of our estimators through an indirect approach, which can establish an approximate upper error bound for our estimator. 

Consider a baseline model that does \textit{not} exploit the smoothness assumption of dynamic networks. Specifically, this baseline model treats all observed networks for $s \in [K], t \in [m]$ as independent layers, without a smoothness assumption on $\mathbf{R}^{[t]}_s$ over time $t$ (we provide more details on this baseline model and its MLE in Appendix~\ref{sec:no-smooth}). 
Note that the difference between the COSIE model and this baseline model is that, the COSIE model defines the probability matrices as $\mathbf{P}^{[t]}_s=\mathbf{X}\mathbf{R}^{[t]}_s\mathbf{Y}^\top$, while this baseline model defines the probability matrices $\mathbf{P}^{[t]}_s$ using a logistic link function as $\text{logit}(\mathbf{P}^{[t]}_s)=\tilde{\mathbf{P}}^{[t]}_s=\mathbf{X}\mathbf{R}^{[t]}_s\mathbf{Y}^\top$. 
The difference between this baseline model and the MFTDN model is that, this baseline model treats all temporal layers as independent layers, while the MFTDN model incorporates a smoothness assumption on $\mathbf{R}^{[t]}_s$ over time $t$.

Intuitively, when the smoothness assumption holds, the MFTDN model should achieve more accurate estimation than this baseline model by leveraging temporal smoothness information. 
(In the revised manuscript, we have added a comparison between MFTDN and this baseline model in the simulation experiments, which confirms that MFTDN indeed achieves better accuracy by exploiting smoothness; see our response to the next comment below for more details.)
And for the MLE of this baseline model, we can derive its upper error bounds, using the asymptotic theory of MLEs. 
%{\color{orange}(RZ: if you are familiar with MLE theory techniques, please add some details about the ``asymptotic theory of MLEs" here)}
Thus, the upper error bounds for the MLE of this baseline model also provide an approximate upper bound on the estimation error of our MFTDN.

For example, if we assume:
\begin{itemize}
    \item $\|\mathbf{X}\|_{2\to\infty}, \|\mathbf{Y}\|_{2\to\infty} \lesssim n^{-1/2}$, where the $2\to\infty$ norm is the maximum row norm;
    \item $\|\mathbf{R}^{[t]}_s\| \asymp n\gamma_n$ for any $t\in\mathcal{T}, s\in[K]$, where $\gamma_n$ is a signal strength parameter, and $\max_{s,t}\sigma_1(\mathbf{R}^{[t]}_s)/\sigma_d(\mathbf{R}^{[t]}_s) \leq C$ for some constant $C$;
    \item $\min_{k,i,j}\mathbf{P}^{[t]}_{k,i,j}(1-\mathbf{P}^{[t]}_{k,i,j}) \geq \rho_n$, where $\rho_n$ is a network density parameter.
\end{itemize}
Under appropriate conditions on $\rho_n$ and $\gamma_n$, applying standard MLE theory to the baseline model, we have that (ignoring logarithmic factors),
\begin{align*}
\|\hat{\mathbf{X}}\mathbf{W}_{\mathbf{X}} - \mathbf{X}\|_{2\to\infty} &\lesssim (Km\rho_n)^{-1/2}(n\gamma_n)^{-1}, \\
\|\hat{\mathbf{Y}}\mathbf{W}_{\mathbf{Y}} - \mathbf{Y}\|_{2\to\infty}&\lesssim (Km\rho_n)^{-1/2}(n\gamma_n)^{-1},\\
\|\mathbf{W}_\mathbf{X}^\top\hat{\mathbf{R}}^{[t]}_k\mathbf{W}_\mathbf{Y} - \mathbf{R}^{[t]}_k\|_F/ \|\mathbf{R}^{[t]}_k\|_F
&\lesssim (\rho_n)^{-1/2}(n\gamma_n)^{-1}
\end{align*}
with high probability,
%for each $i\in[n]$, where $\mathbf{x}_i$, $\hat{\mathbf{x}}_i$, $\mathbf{y}_i$, $\hat{\mathbf{y}}_i$ denote the $i$-th row of $\mathbf{X}$, $\hat{\mathbf{X}}$, $\mathbf{Y}$, and $\hat{\mathbf{Y}}$, respectively, and 
where $\mathbf{W}_\mathbf{X}=\arg\min_{\mathbf{O}\in\mathcal{O}_d}\|\hat{\mathbf{X}}\mathbf{O} -\mathbf{X}\|_F$ and $\mathbf{W}_\mathbf{Y}=\arg\min_{\mathbf{O}\in\mathcal{O}_d}\|\hat{\mathbf{Y}}\mathbf{O} -\mathbf{Y}\|_F$. 
These bounds provide an approximate upper error bound for our MFTDN estimators.

}

\newpage

\section{Proofs of Stated Results}

\subsection{Proof of Proposition~\ref{prop:identifiability}}

{\color{black}
%Proposition~\ref{prop:identifiability} follows directly from Proposition~2 in \citet{arroyo2021inference}.

This proof adapts the arguments from Proposition~2 in \cite{arroyo2021inference}.

Since $(\mathbf{X},\mathbf{Y},\mathbf{R}^{[t]}_s)$ and $(\mathbf{X}',\mathbf{Y}',(\mathbf{R}')^{[t]}_s)$ yield the same probability matrices $\{\mathbf{P}^{[t]}_s\}_{t\in\mathcal{T},s\in[K]}$, we have
\begin{equation}\label{eq:XRY=XRY}
	\mathbf{X}'(\mathbf{R}')^{[t]}_s\mathbf{Y}'^{\top}
	=\mathbf{X}\mathbf{R}^{[t]}_s\mathbf{Y}^\top
\quad\text{for all }t\in\mathcal{T},s\in[K].
\end{equation}
By \eqref{eq:XRY=XRY} and the orthonormality of $\mathbf{Y}'$, $\mathbf{Y}$, and $\mathbf{X}$, we have
\begin{align}
\sum_{t\in\mathcal{T},s\in[K]}\mathbf{X}'(\mathbf{R}')^{[t]}_s\mathbf{Y}'^\top\mathbf{Y}'(\mathbf{R}')^{[t]\top}_s\mathbf{X}'^\top
&=\sum_{t\in\mathcal{T},s\in[K]}\mathbf{X}\mathbf{R}^{[t]}_s\mathbf{Y}^\top\mathbf{Y}\mathbf{R}^{[t]\top}_s\mathbf{X}^\top \notag\\
\mathbf{X}'\left(\sum_{t\in\mathcal{T},s\in[K]}(\mathbf{R}')^{[t]}_s(\mathbf{R}')^{[t]\top}_s\right)\mathbf{X}'^\top
&=\mathbf{X}\left(\sum_{t\in\mathcal{T},s\in[K]}\mathbf{R}^{[t]}_s\mathbf{R}^{[t]\top}_s\right)\mathbf{X}^{\top} \notag\\
\mathbf{X}'\big(\tilde{\mathbf{R}}'\tilde{\mathbf{R}}'^\top\big)\mathbf{X}'^\top
&=\mathbf{X}\big(\tilde{\mathbf{R}}\tilde{\mathbf{R}}^\top\big)\mathbf{X}^{\top} \notag\\
\mathbf{X}'\big(\tilde{\mathbf{R}}'\tilde{\mathbf{R}}'^\top\big)\mathbf{X}'^\top\mathbf{X}
&=\mathbf{X}\big(\tilde{\mathbf{R}}\tilde{\mathbf{R}}^\top\big)\mathbf{X}^{\top}\mathbf{X} \notag\\
\mathbf{X}'\big(\tilde{\mathbf{R}}'\tilde{\mathbf{R}}'^\top\big)\mathbf{X}'^\top\mathbf{X}
&=\mathbf{X}\big(\tilde{\mathbf{R}}\tilde{\mathbf{R}}^\top\big), \label{eq:XRRXX}
\end{align}
where we define
$
	\tilde{\mathbf{R}}':=\left[ ({\mathbf{R}'})_1^{[t_1]}, ({\mathbf{R}'})_1^{[t_2]}, \dots, ({\mathbf{R}'})_K^{[t_m]} \right],
	\tilde{\mathbf{R}}:=\left[ {\mathbf{R}}_1^{[t_1]}, {\mathbf{R}}_1^{[t_2]}, \dots, {\mathbf{R}}_K^{[t_m]} \right].
$
Note that the full-rank condition on $\tilde{\mathbf{R}}$ implies that $\tilde{\mathbf{R}}\tilde{\mathbf{R}}^\top$ is also full-rank, so the inverse of $\tilde{\mathbf{R}}\tilde{\mathbf{R}}^\top$ exists.
Then, by \eqref{eq:XRRXX}, we have
\begin{equation}\label{eq:XWX}
\mathbf{X}'\tilde{\mathbf{W}}_{\mathbf{X}}
=\mathbf{X},
\end{equation}
where we define
$
\tilde{\mathbf{W}}_{\mathbf{X}}:=\big(\tilde{\mathbf{R}}'\tilde{\mathbf{R}}'^\top\big)\mathbf{X}'^\top\mathbf{X}\big(\tilde{\mathbf{R}}\tilde{\mathbf{R}}^\top\big)^{-1}.
$
We next verify that $\tilde{\mathbf{W}}_{\mathbf{X}}$ is orthogonal.
By \eqref{eq:XWX}, we have $\tilde{\mathbf{W}}_{\mathbf{X}}=\mathbf{X}'^\top\mathbf{X}'\tilde{\mathbf{W}}_{\mathbf{X}}=\mathbf{X}'^\top\mathbf{X}$
and $\mathbf{X}^\top\mathbf{X}'\tilde{\mathbf{W}}_{\mathbf{X}}=\mathbf{X}^\top\mathbf{X}=\mathbf{I}$, which imply $\tilde{\mathbf{W}}_{\mathbf{X}}^\top\tilde{\mathbf{W}}_{\mathbf{X}}=\mathbf{I}$.
Since $\tilde{\mathbf{W}}_{\mathbf{X}}$ is a square matrix of size $d\times d$, the uniqueness of the matrix inverse guarantees that $\tilde{\mathbf{W}}_{\mathbf{X}}^{-1}=\tilde{\mathbf{W}}_{\mathbf{X}}^\top$, and thus $\tilde{\mathbf{W}}_{\mathbf{X}}$ is orthogonal.

We define ${\mathbf{W}}_{\mathbf{X}}:=\tilde{\mathbf{W}}_{\mathbf{X}}^\top$. Then ${\mathbf{W}}_{\mathbf{X}}$ is also orthogonal, and by \eqref{eq:XWX} we have
\begin{equation}\label{eq:XWX2}
\mathbf{X}{\mathbf{W}}_{\mathbf{X}}
=\mathbf{X}'.
\end{equation}
Similarly, there exists an orthogonal matrix $\mathbf{W}_{\mathbf{Y}}$ such that 
\begin{equation}\label{eq:YWY}
\mathbf{Y}\mathbf{W}_{\mathbf{Y}}
=\mathbf{Y}'.
\end{equation}
Finally, for any $t\in\mathcal{T}, s\in[K]$, by \eqref{eq:XRY=XRY}, \eqref{eq:XWX2}, \eqref{eq:YWY}, and the orthonormality of $\mathbf{X}$ and $\mathbf{Y}$, we have
\begin{align*}
\mathbf{X}\mathbf{W}_{\mathbf{X}}(\mathbf{R}')^{[t]}_s\mathbf{W}_{\mathbf{Y}}^\top\mathbf{Y}^\top
&= \mathbf{X}\mathbf{R}^{[t]}_s\mathbf{Y}^\top \\
\mathbf{X}^\top\mathbf{X}\mathbf{W}_{\mathbf{X}}(\mathbf{R}')^{[t]}_s\mathbf{W}_{\mathbf{Y}}^\top\mathbf{Y}^\top\mathbf{Y}
&= \mathbf{X}^\top\mathbf{X}\mathbf{R}^{[t]}_s \mathbf{Y}^\top\mathbf{Y}\\
\mathbf{W}_{\mathbf{X}}(\mathbf{R}')^{[t]}_s\mathbf{W}_{\mathbf{Y}}^\top
&= \mathbf{R}^{[t]}_s\\
(\mathbf{R}')^{[t]}_s
&= \mathbf{W}_{\mathbf{X}}^\top\mathbf{R}^{[t]}_s\mathbf{W}_{\mathbf{Y}}.
\end{align*}
}

\subsection{Proof of Proposition~\ref{proposition_representer_thm}}

    For any fixed $\mx$ and $\my$ with {\red column norm $1$}, by the Karush-Kuhn-Tucker optimality condition in Hilbert Space (e.g., Theorem 5.1 in chapter 3 of \cite{ekeland1999convex}) and the convexity of the loss function, there exists some constant $\lambda$ such that
    the constrained optimization problem 
    \bea\label{eq_tmp1}
     \red\argmin_{{\substack{\mr_{s, k,\ell} \in \mathcal{H} \\ \sum_{s \in [K]} \sum_{k,l\in[d]} \|\mr_{s, k,\ell}\|_{\mathcal{H}} \leq C}}} 
     \sum_{s \in [K]}\sum_{t \in \mathcal{T}} \sum_{i,j\in[n]} f\(\ma^{[t]}_{s,i,j}, \sum_{\ell,k\in[d]}(\mathbf{x}_i)_k \, \mr^{[t]}_{s, k,\ell} \, (\mathbf{y}_j)_\ell\)
    \eea can be replaced by the following regularized problem:
    \[
        \argmin_{\mr_{s, k,\ell} \in \mathcal{H}; s\in[K],\ell, k \in [d]} 
        \sum_{s \in [K]}\sum_{t \in \mathcal{T}} \sum_{i,j\in[n]}
        f\(\ma^{[t]}_{s,i,j}, \sum_{\ell,k\in[d]}(\mathbf{x}_i)_k \, \mr^{[t]}_{s, k,\ell} \, (\mathbf{y}_j)_\ell\)+ \lambda^\prime \|\mr\|_\mathcal{H},
    \]
    where $\|\mr\|_\mathcal{H}^2 := \sum_{s \in [K]}\sum_{\ell,k\in[d]} \|\mr_{s, k,\ell}\|_{\mathcal{H}}^2$. 
   Note that in \eqref{eq_tmp1}, the feasible set is bounded and closed, which is also weakly compact by Banach-Alaoglu Theorem \citep{rudin1991functional}. Further note that the loss function is Gateaux-differentiable and convex on $\mr^{[t]}_{s, k,\ell}$. We claim that a convex Gateaux differentiable map on $\mathcal{H}$ attains its minimum when restricted to any weakly compact subset. 
    Furthermore, any minimizer to the regularized problem admits a form of $\hat \mr^{[t]}_{s, k,\ell} = \sum_{h \in \mathcal{T}} \theta_{s, k,\ell, h} \mathbb{K}(t, h)$ by a generalized form of the Representer Theorem \citep{scholkopf2001generalized}. 

    Finally, we prove the claim. 
    Let $K \subseteq \mathcal{H}$ be a weakly compact subset. Consider a minimizing sequence $\{x_n\} \subset K$ such that $F(x_n) \to \inf_{x \in K} F(x)$. 
    Since $K$ is weakly compact, there exists a subsequence and an element $x^* \in K$ such that $x_{n_i} \to x^*$ weakly in $E$. {\red By convexity of $F$, we have:
    \[
    F(x_{n_i}) \ge F(x^*) + \left\langle \nabla F(x^*),\, x_{n_i} - x^* \right\rangle .
    \]
    }
    Let $n \to \infty$. By the weak continuity of the directional derivative, it follows $\liminf_{n \to \infty} F(x_n) \geq F(x^*)$. 
    Thus it follows that $\inf_{x \in K} F(x) \geq F(x^*)$, and hence $
    F(x^*) = \inf_{x \in K} F(x)$. 

{\color{black}

\newpage

\section{Additional Experiment Results and Discussion}

\subsection{A baseline model for Section~\ref{sec:multi}}
\label{sec:no-smooth}

As a baseline for the experiments in Section~\ref{sec:multi}, we describe the algorithm for a version of MFTDN without temporal smoothing. That is, we treat all observed networks for $s \in [K],t \in [m]$ as independent layers, without imposing smoothness constraints on $\mathbf{R}^{[t]}_s$ over time $t$.

Given the observed adjacency matrices $\{\mathbf{A}^{(s)}\}_{s=1}^m$, the log-likelihood function is
$$
\ell(\mathbf{X}, \{\mathbf{R}^{[t]}_s\}, \mathbf{Y}) =\sum_{s\in[K],t\in\mathcal{T},i\in[n],j\in[n]} \left[ \mathbf{A}^{[t]}_{s,i,j} \log \mathbf{P}^{[t]}_{s,i,j} + (1-\mathbf{A}^{[t]}_{s,i,j}) \log(1-\mathbf{P}^{[t]}_{s,i,j}) \right]
$$
where $\operatorname{logit}(\mathbf{P}^{[t]}_s) = \tilde{\mathbf{P}}^{[t]}_s=\mathbf{X}\mathbf{R}^{[t]}_s\mathbf{Y}^\top$.
With $\mathbf{E}^{[t]}_s = \mathbf{A}^{[t]}_s - \mathbf{P}^{[t]}_s$, the gradients are
$$
\frac{\partial \ell}{\partial \mathbf{R}^{[t]}_s} = \mathbf{X}^\top \mathbf{E}^{[t]}_s \mathbf{Y}, \quad \frac{\partial \ell}{\partial \mathbf{X}} = \sum_{s\in[K],t\in\mathcal{T}} \mathbf{E}^{[t]}_s \mathbf{Y} (\mathbf{R}^{[t]}_s)^\top, \quad \frac{\partial \ell}{\partial \mathbf{Y}} = \sum_{s\in[K],t\in\mathcal{T}} (\mathbf{E}^{[t]}_s)^\top \mathbf{X} \mathbf{R}^{[t]}_s.
$$
We estimate the parameters via alternating gradient ascent, which iteratively updates $\{\mathbf{R}^{[t]}_s\}$, $\mathbf{X}$, and $\mathbf{Y}$ by ascending their respective gradients while holding the other parameters fixed.

{\color{black}

\subsection{Additional evaluation of $\mathbf R$ at observed time points}
\label{app:obs_R_accuracy}

In Section \ref{sec:simu_error}, the accuracy of the estimated functional matrices $\hat{\mathbf R}^{[t]}_s$ is evaluated on a dense grid of time points to assess how well the fitted RKHS trajectory recovers the underlying dynamic pattern. Since this evaluation includes time points that are not observed in the training data, we further report the accuracy of $\hat{\mathbf R}^{[t]}_s$ restricted to the observed design points $\mathcal T=\{t_1,\ldots,t_m\}$.

Specifically, after aligning the estimated latent subspaces, we compute
\[
    \operatorname{Acc}_{\mathrm{obs}}(\hat{\mathbf R})
    =
    \operatorname{Corr}\left(
    \mathbf W_{\mathbf X}^{\top}\left[
    \hat{\mathbf R}^{[t_1]}_1
    \mid \cdots \mid
    \hat{\mathbf R}^{[t_m]}_K 
    \right]\mathbf W_{\mathbf Y},
    \left[
    \mathbf R^{[t_1]}_1
    \mid \cdots \mid
    \mathbf R^{[t_m]}_K
    \right]
    \right),
\]
where $\mathbf W_{\mathbf X}$ and $\mathbf W_{\mathbf Y}$ are the alignment matrices used in the main simulation study. Figure~\ref{fig:obs_R_accuracy} reports this observed-design-point accuracy under the same settings as Figure~\ref{fig:simu_error}.
\begin{figure}[htbp]
    \centering
    \includegraphics[width=0.95\textwidth]{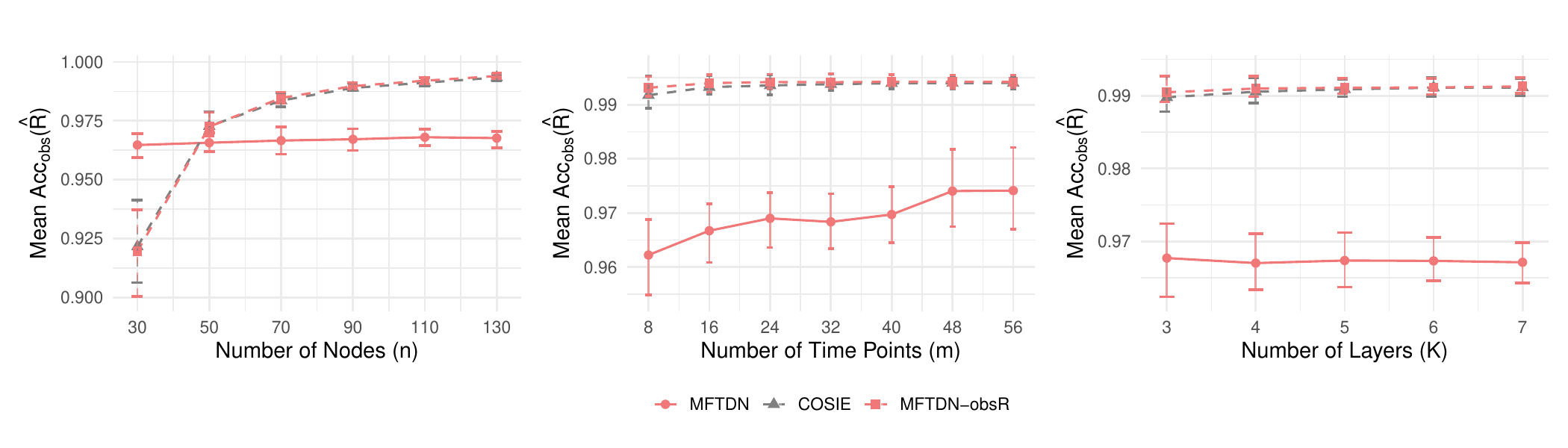}
    \caption{Observed-design-point accuracy of $\hat{\mathbf R}^{[t]}_s$ for MFTDN. The accuracy is computed only at the observed time points $\mathcal T=\{t_1,\ldots,t_m\}$ under the same simulation settings as Figure~\ref{fig:simu_error}. }
    \label{fig:obs_R_accuracy}
\end{figure}

The results show that the accuracy of $\hat{\mathbf{R}}^{[t]}_s$ at the observed design points is close to its accuracy evaluated on the dense time grid. This suggests that the fitted RKHS trajectory is faithful to the underlying dynamic pattern, rather than merely producing smooth interpolation between observed time points.

We also observe that the observed-design-point accuracy of MFTDN is slightly lower than that of the COSIE baseline. This is expected because the no-smoothing baseline estimates each $\mathbf{R}_s^{[t]}$ independently at each observed time point and is therefore directly optimized for design-point fit. In contrast, MFTDN estimates a single functional trajectory through the RKHS representation, which is designed to recover the smooth temporal evolution of $\mathbf{R}_s^{[t]}$ over the entire time domain rather than to maximize pointwise accuracy at each observed time point. Thus, this comparison reflects a tradeoff: the no-smoothing baseline can fit the observed design points more closely, while MFTDN provides a coherent smooth functional estimate that generalizes to unobserved time points. 

Two important points merit emphasis. First, the slightly lower accuracy at observed design points does not imply that MFTDN produces less accurate estimates of $\mathbf{R}_s^{[t]}$ overall. On the contrary, MFTDN can yield a more accurate global estimate of $\mathbf{R}_s^{[t]}$ across the entire time domain. This is evidenced in Section~\ref{sec:multi}, where the clustering accuracy derived from the estimated $\mathbf{R}_s^{[t]}$ demonstrates that MFTDN outperforms the no-smoothing baseline. Second, if we fully leverage the information observed at the design points after obtaining $\hat{\mathbf{X}}$ and $\hat{\mathbf{Y}}$ from MFTDN by estimating $\mathbf{R}_s^{[t]}$ at observed points via $\hat{\mathbf{R}}_s^{[t]} = \hat{\mathbf{X}}^\top \mathbf{A}_s^{[t]} \hat{\mathbf{Y}}$ (as in COSIE), we obtain the dashed line labeled MFTDN-obsR in Figure~\ref{fig:obs_R_accuracy}. As shown, the accuracy of MFTDN-obsR is slightly better than that of COSIE.

%This is expected because the no-smoothing baseline estimates each $\mathbf R_s^{[t]}$ separately at each observed time point and is therefore directly optimized for design-point fit. In contrast, MFTDN estimates a single functional trajectory through the RKHS representation, which is designed to recover the smooth temporal evolution of $\mathbf R_s^{[t]}$ over time rather than to maximize pointwise accuracy at each observed time point. Thus, the comparison should be interpreted as showing a tradeoff: the no-smoothing baseline can fit the observed design points more closely, while MFTDN provides a coherent smooth functional estimate that also generalizes to unobserved time points.
%Moreover, as can be seen from Figure~\ref{app:obs_R_accuracy}, as $m$ increases, the accuracy of $\mathbf R_s^{[t]}$ estimates from the COSIE baseline does not improve, while MFTDN obtains improved estimates of the $\mathbf R_s^{[t]}$ matrices and the gap between MFTDN and the no-smoothing baseline decreases.

}

\subsection{Additional results for Section~\ref{sec:citi}}\label{sec:citi_add}

\begin{figure}[htbp!]
    \centering
    \includegraphics[width=0.75\textwidth]{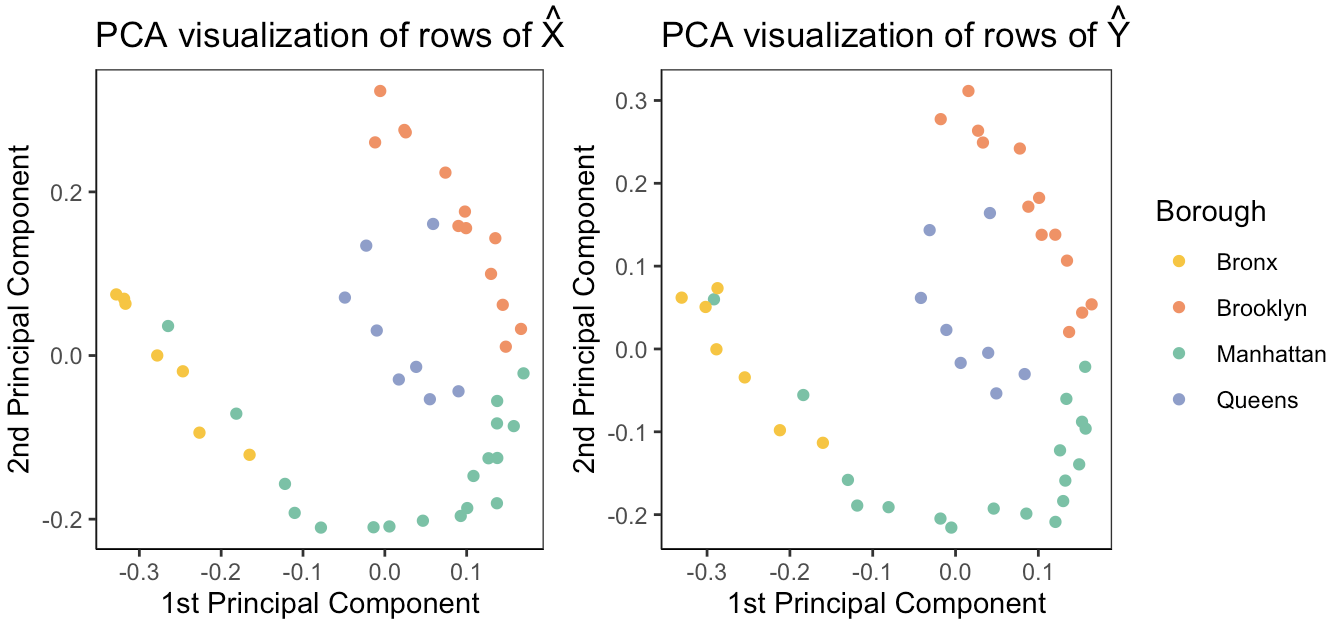}
    \caption{PCA visualization of the rows of $\hat{\mx}$ and $\hat{\my}$ using the first two principal components, with colors indicating the boroughs of the vertices.}
    \label{fig:real_citi_XY_PCA}
\end{figure}

\subsection{Additional results for Section~\ref{sec:trade}}\label{sec:trade_add}

Regarding the choice of kernel and embedding dimension, we compare the BIC for Bernoulli, polynomial, radial, and periodic kernels across different embedding dimensions $d$. Figure~\ref{fig:real_trade_BIC} shows that the periodic kernel achieves the best performance, with comparable BIC values at $d=3$ and $d=4$. For model parsimony, we select the periodic kernel with $d=3$.

\begin{figure}[htbp]
    \centering
    \includegraphics[width=0.5\textwidth]{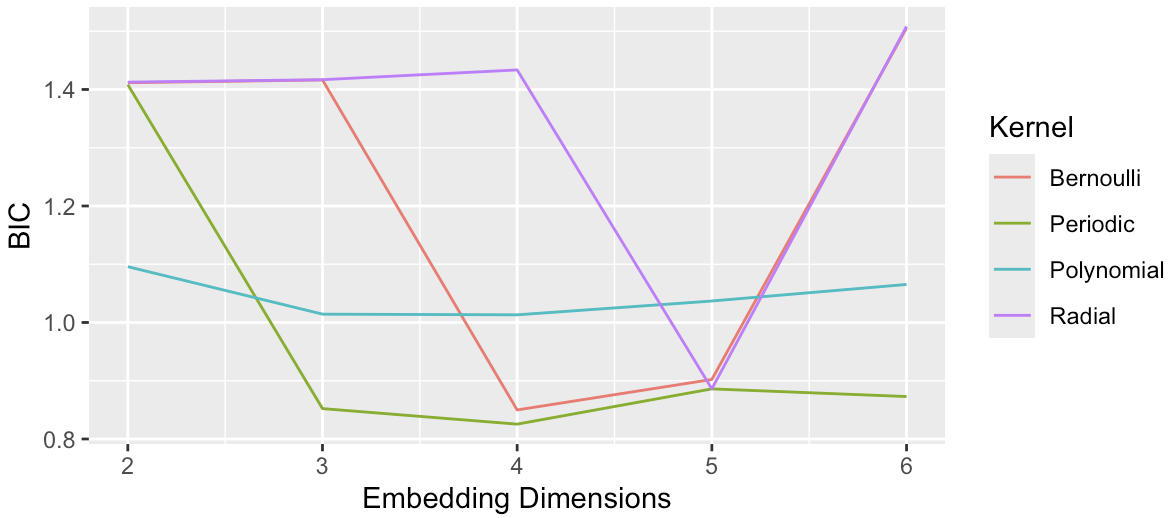}
    \caption{Kernel and embedding dimension selection by BIC.}
    \label{fig:real_trade_BIC}
\end{figure}

\begin{figure}[htbp]
    \centering
    \includegraphics[width=0.9\textwidth]{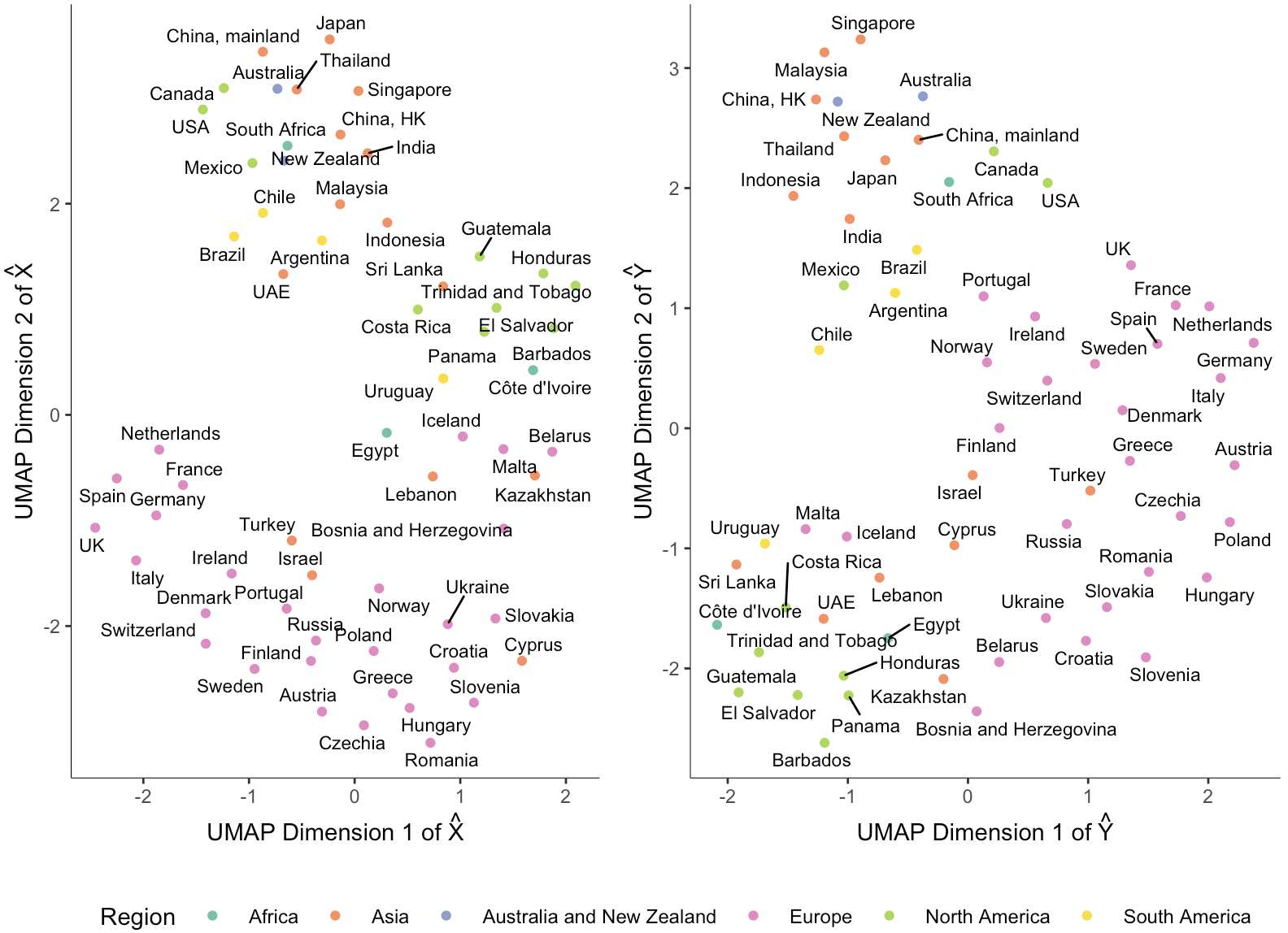}
    \caption{UMAP visualization of the embeddings of trade entities from $\hat{\mx}$ and $\hat{\my}$.}
    \label{fig:real_trade_XY_UMAP}
\end{figure}

}

\end{appendices}

\end{sloppypar}

\end{document}